# A lab-on-a-silicon-chip platform for all-electrical antibiotic susceptibility tests with a sample-to-results time within 20 minutes

Zheqiang Xu,[1] Victoria C. Nolan,[1] Yingtao Yu,[1] Petra Muir,[2] Sanna Koskiniemi, [2*] and Zhen Zhang[1]*

[1]Ångström Laboratory, Division of Solid-State Electronics, Department of Electrical Engineering, Uppsala University, SE-75121 Uppsala, Sweden.

[2]Department of Cellular and Molecular Biology, Uppsala University, SE-75124 Uppsala, Sweden

* Corresponding author. Email: *sanna.koskiniemi@icm.uu.se*; *zhen.zhang@angstrom.uu.se*

**Abstract:** Rapid antibiotic susceptibility tests (ASTs) are essential for quick selection of effective drugs to treat infections at an early stage. However, the most widely used phenotypic ASTs in clinical practice often require 24–48 hours of pre-culture enrichment and 4–20 hours of testing. They are too slow to guide therapy, even with the most rapid protocol. Here, we report a lab-on-a-silicon chip (LOSC) system, which integrates arrays of silicon nanowire field-effect transistor (SiNWFET) sensors with high-throughput cell-collection microfluidics for rapid ASTs. The microfluidics concentrate bacteria into picoliter-scale chambers within minutes, eliminating the need for pre-cultivation. Embedded SiNWFETs sensitively track antibiotic-induced metabolic pH shifts, enabling AST by detecting rapid metabolic inhibition in susceptible bacteria. Using an unbuffered culturing medium, LOSC achieves sample-to-result times within 20 minutes for clinically isolated uropathogenic *Escherichia coli* strains. With its electrical readout and compact design, LOSC offers a low-cost, rapid, and portable AST solution for point-of-care diagnostics.

**Teaser:** A lab-on-a-silicon-chip platform offers a rapid and low-cost AST option for urine samples.

## 1. INTRODUCTION

Antibiotic/antimicrobial resistance (AMR) poses a critical threat to global health (*1, 2*), demanding rapid and accurate diagnostic tools to guide effective antibiotic therapy (*3, 4*). Phenotypic antibiotic susceptibility tests (ASTs) can determine appropriate antibiotic treatments for bacterial infections, thus ensuring the use of effective antibiotics and preventing the spread of AMR (*5-9*). However, the most widely used phenotypic tests in clinical practice, such as broth microdilution and Kirby-Bauer disk diffusion (*10*), are culture-based and time-consuming, requiring 24-48 hours of pre-cultivation before the tests can start. Even with the most recent rapid AST protocol (*11*), the traditional test still takes 8-20 hours, which is too long for patients to wait for therapy. Genotypic AST, detecting specific genetic sequences associated with antibiotic resistance (*12, 13*), can provide a more rapid test (1–3 hours) (*14, 15*), but cannot detect previously unidentified resistance mechanisms (false negative) (*16*). In addition, genotypic tests can also give false positives, since the presence of a resistance gene is not always associated with actual resistance (*17*).

Over the past decades, numerous innovative phenotypic ASTs have been developed (*7*), including motion detection (*18, 19*), impedance-based growth monitoring (*20*), electrochemical sensing (*21, 22*), Raman-assisted metabolic method (*23-25*), and microfluidic-based optical methods (*17, 26-28*). Among these, a promising strategy for rapid AST involves monitoring bacterial metabolism under antibiotic treatment (*29-31*). Since antibiotics alter metabolic processes before causing cell morphological changes or cell death (*32-34*), metabolic profiling can potentially be faster than monitoring the cell morphology or cell numbers to determine antibiotic efficacy.

We previously demonstrated bacteria metabolism monitoring using silicon nanowire field-effect transistors (SiNWFETs) (*35*). The SiNWFET sensors electrically monitor cell metabolism-induced pH changes in the culture media, profiling detailed metabolic responses of bacteria to different antibiotics. It achieved a rapid testing time within 30 minutes. However, this method requires a large volume of samples with high cell density (~$10^9$ cells/mL) to enable rapid pH changes, which necessitates a pre-cultivation step for clinical samples and extends the sample-to-result time.

To address this limitation, we further integrated SiNWFET sensor arrays with high-throughput cell-collection microfluidics to realize a lab-on-a-silicon chip (LOSC) system for rapid ASTs without the need for pre-cultivation. *Escherichia coli* (*E. coli*) was chosen as a model pathogen in this study because it causes approximately 85% of urinary tract infections (UTIs) in primary care (*17*). With an estimated 450 million UTI cases globally each year (*36*) and a high frequency of resistance to first-line antibiotics (*37, 38*), the need for rapid UTI ASTs is urgent. UTI samples are clinically relevant if they contain $10^3$–$10^5$ cells/mL of *E. coli* (*17, 39*). In our LOSC system, the microfluidics can collect and concentrate *E. coli* from dilute samples into picoliter-scale culture chambers within 10 minutes. The concentration of cells will accelerate pH changes in the culture chambers. In addition, by using an unbuffered medium, the LOSC system can determine the drug effect on clinically isolated *E. coli* within 10 minutes, yielding a total sample-to-result time of under 20 minutes. We further evaluated the LOSC platform across different bacteria, validating both its microfluidic capture of additional uropathogens and its AST performance on the Gram-positive/Gram-negative bacterium. These results demonstrated a rapid electrical AST platform with significantly reduced requirements on sample preparation, which holds great promise for low-cost point-of-care applications.

# 2. RESULTS

## 2.1 The LOSC design and realization

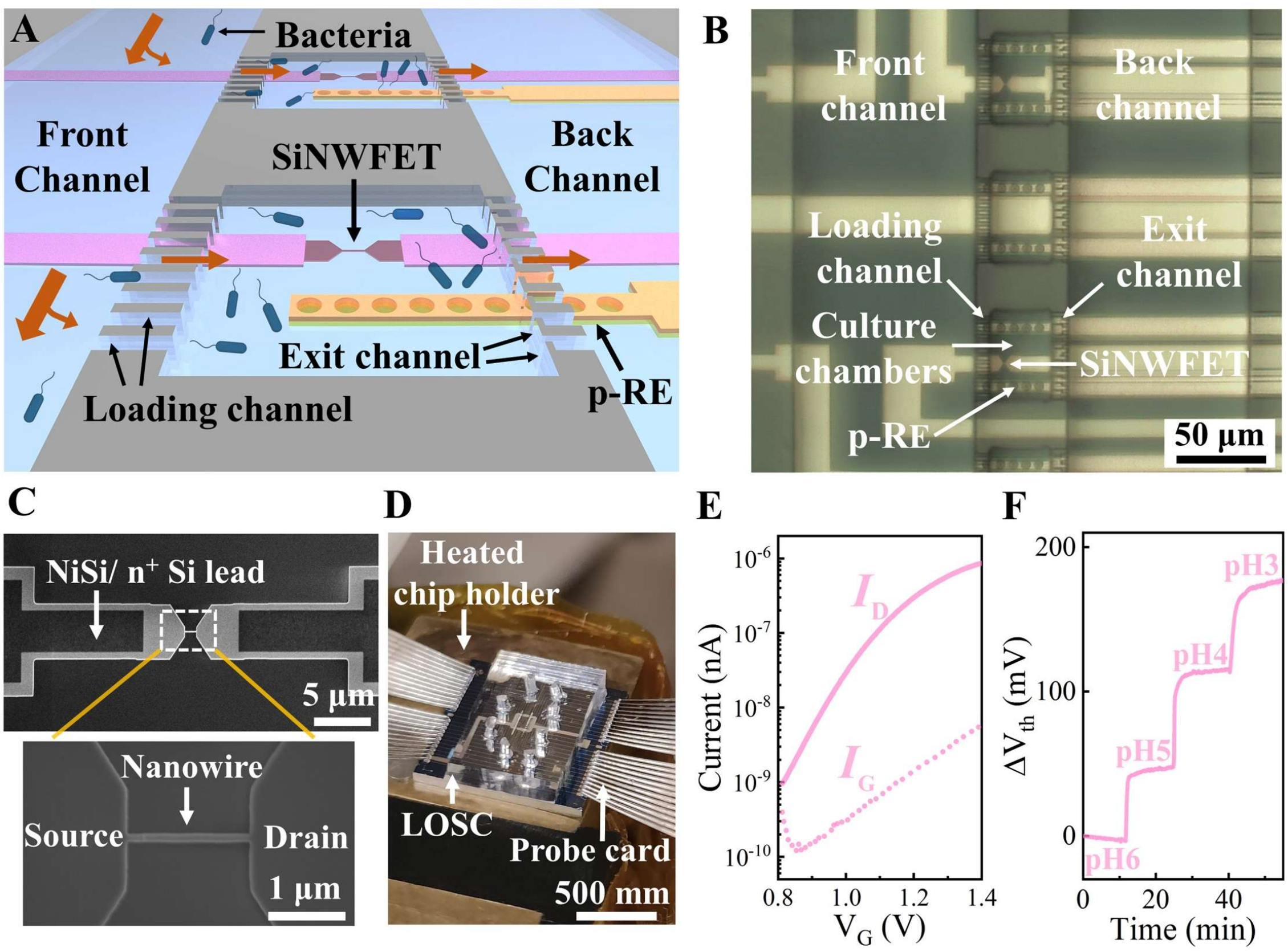

**Fig.1. The design of the LOSC.** (**A**) Schematic illustration of the LOSC integrating a silicon sensor chip with cell-collection microfluidics. In microfluidics, microscale culture chambers are positioned between a front channel and a back channel, connected via loading and exit channels, respectively. Each culture chamber houses a SiNWFET sensor paired with an on-chip AgCl/Ag pseudo reference electrode (p-RE) on the silicon sensor chip. (**B**) Top-view optical image of the integrated LOSC. (**C**) Top-view scanning electron microscope (SEM) image of a SiNWFET. (**D**) Optical image of the measurement setup of the LOSC. (**E**) Transfer characteristics ($I_D$-$V_G$) of a SiNWFET measured at $V_{DS}$ = 0.3 V in pH 7, 0.1 M NaCl at 37 °C, with gate bias applied via the on-chip AgCl/Ag p-RE. (**F**) pH responses of the SiNWFET sensor in 0.1 M NaCl at 37 °C. The shaded area along the solid lines depicts the standard deviations calculated from 3 SiNWFET sensors.

The LOSC integrates a silicon sensor chip containing arrays of SiNWFET sensors, with cell-collection microfluidics, as illustrated in Fig. 1(A and B) (More details on the LOSC fabrication can be found in *Materials and methods* and Fig. S1-S4.). In microfluidics, arrays of microscale culture chambers are integrated between a front channel and a back channel, with connections via sieve-like loading channels and exit channels, respectively (Fig. 1A). The loading channels have a cross-sectional dimension of 2 μm in width and 1.5 μm in height, which permits the passage of *E. coli*. While the exit channels feature a minimum width of 0.5 μm to block *E. coli* escape with the liquid flow (Fig. 1B). Each culture chamber has a tiny volume of 35 pL (50×35×20 μm) and houses a SiNWFET sensor, together with the pairing AgCl/Ag pseudo reference electrode (p-RE) (*40*). The nanowire channel of the SiNWFET is 100 nm wide and 1.6 μm long (Fig. 1C).

To perform ASTs, the LOSC was connected to readout (HP 4155) via a probe card (Fig. 1D). We first characterized the SiNWFET sensor's pH response by applying gate voltage ($V_G$) through the on-chip AgCl/Ag p-RE. The transfer curve ($I_D$ - $V_G$) in 0.1 M NaCl solution at pH 7 shows n-type behavior, a subthreshold swing of 120 mV/dec, $I_{on}/I_{off}$ of $10^3$, and gate leakage below 3 nA (Fig. 1E). For pH sensing, a constant $V_G$ was applied, and $I_D$ was recorded versus time. The shift of the threshold voltage of the SiNWFET ($\Delta V_{th}$), induced by adsorption of protons to the pH sensing layer on the sensor surface (*35*), was extracted from the $I_D$ change using the $I_D$-$V_G$ curve (*40*). We observed a near-Nernstian $\Delta V_{th}$ response to pH with a sensitivity of 59.4 ± 0.8 mV/pH at 37 °C (Fig. 1F). The noise of the SiNWFET sensor was evaluated from the power spectral density (PSD) of $V_G$ (Fig. S5). The root mean square (RMS) noise of $V_G$ is below 1 mV, indicating that the device noise has a negligible influence on metabolic signal detection.

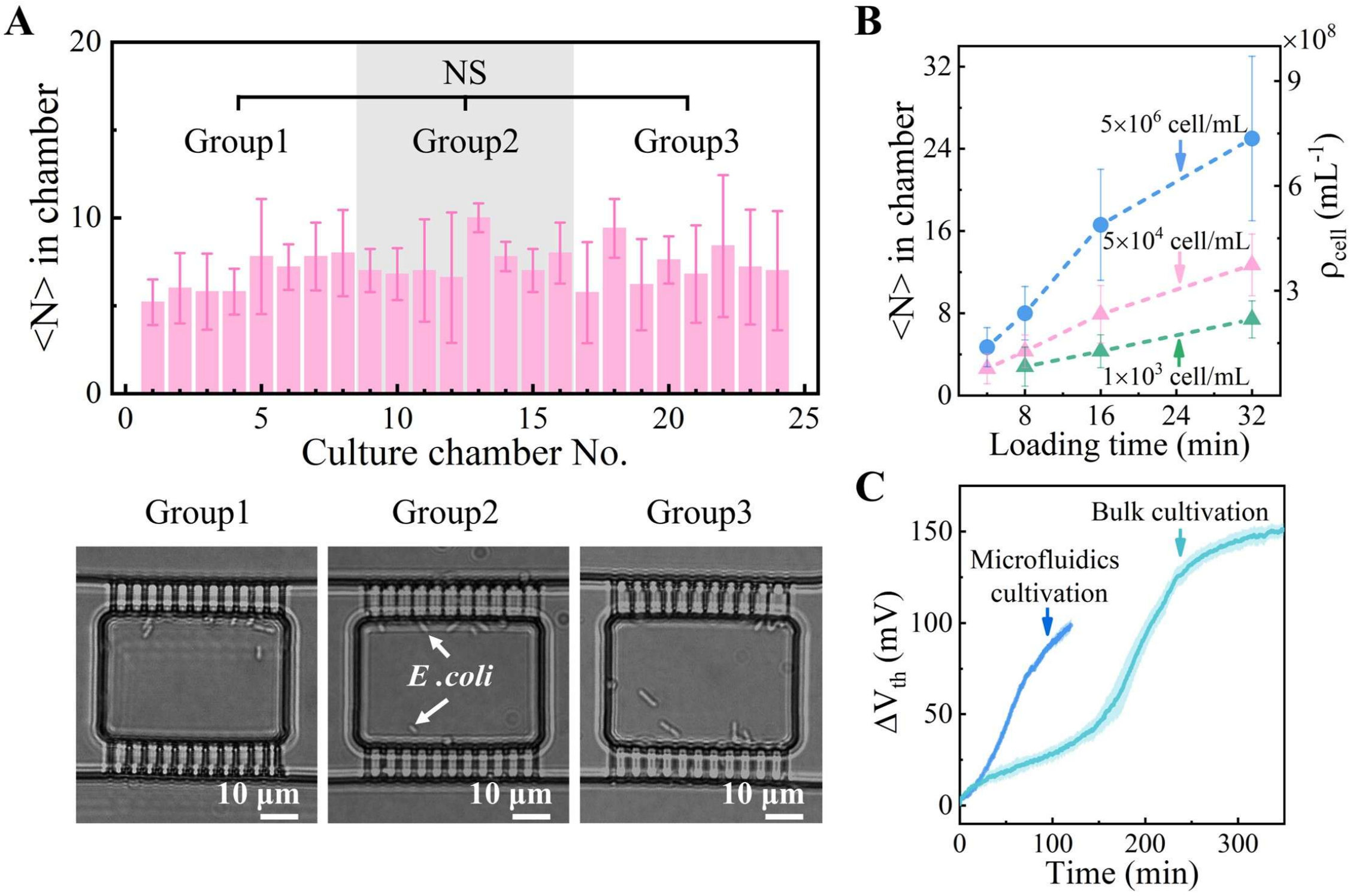

**Fig. 2. *E. coli* capture in cell-collection microfluidics.** (**A**) *E. coli* distribution across 24 culture chambers after 8-minute bypass-loading using a sample with $5\times10^6$ cells/mL at 8 μL/min. Bar plots show average cell numbers in chambers; error bars indicate standard deviations from four independent experiments. Chambers are sequentially numbered from inlet (No.1) to outlet (No.24) and grouped into three position-based groups. A one-way ANOVA test showed no significant (NS) differences between the groups ($p > 0.05$). Representative optical images from each group are shown below the bar plot. (**B**) Cell loading performance using samples with varying cell density. Force-loading mode was used for $1\times10^3$ (green) and $5\times10^4$ (pink) cells/mL; Bypass-loading was used for $5\times10^6$ (blue) cells/mL. Average number of cells per chamber (<

$N >$-left axis) and corresponding effective volumetric cell density ($\rho_{cell}$ - right axis) are plotted versus loading times (flow rate 8 μL/min). Effective volumetric cell density was calculated as $\rho_{cell} =< N >/V_{chamb}$ . Error bars represent standard deviations across 24 chambers. (**C**) $\Delta V_{\text{th}}$ vs. time curve for *E. coli* cultivated in microfluidics chambers after 8-minute bypass-loading of $5\times10^6$ cells/mL sample and in a 100 μL PDMS container with $5\times10^6$ cells/mL sample (bulk cultivation, See Fig. S11). LB media with 1 wt.% glucose was used. The shaded areas represent standard deviations from three SiNWFET sensors.

The workflow of the LOSC-based AST involves trapping bacterial cells into the culture chambers, introducing test media, and sealing both the front and back channels with oil to isolate the culture chambers (Fig. S6 for details). This workflow ensures that ASTs are performed in a standardized medium independent of the original sample matrix, improving consistency and comparability. During bacterial trapping, for extremely dilute samples ($<10^5$ cells/mL), we employed a force-loading mode of the microfluidics to trap *E. coli*. In this mode, the outlet of the front channel is temporarily closed, directing the entire sample flow through the culture chambers before exiting via the back channel. This configuration enhances bacterial capture efficiency by ensuring that all fluid is forced through the chambers. For moderately dilute samples ($>10^5$ cells/mL), a bypass-loading mode was adopted to prevent clogging at the loading channel. In this mode, the outlet of the front channel remains open, allowing the majority of the sample to pass along the main channel, with a small fraction (~4.6% from simulations, Fig. S8A) diverted into the chambers via side pathways.

Using an optical microscope, we observed uniform *E. coli* distribution across culture chambers after bypass-loading (Fig. 2A). One-way ANOVA analysis (See Table. S1) confirmed that there was no significant variation in cell distribution across chambers located at different positions, under either loading mode. Bypass-loading allowed capture of about 8 cells in each chamber in 8 minutes when using $5 \times 10^6$ cells/mL starting culture. (Fig. 2B). Even using more diluted samples with $1 \times 10^3$ and $5 \times 10^4$ cells/mL, around 2 - 4 *E. coli* cells per chamber could be collected after 8 minutes of forced-loading (Fig. 2B).

The uniform cell distribution observed in Fig. 2A and Table. S1 is supported by numerical simulations. Under bypass-loading conditions, the simulated flow fields (Fig. S8B) indicate a self-regulated flow-partitioning effect: once bacteria are trapped at the exit channel, the local hydraulic resistance increases and redistributes the incoming flow toward less occupied chambers, promoting uniform loading. Under force-loading conditions, simulations show a nearly identical flow distribution among chambers (Fig. S9A and B), while the increased pressure drop across the loading channels (Fig. S9D) efficiently forces bacteria into each chamber. Importantly, although force-loading operates at elevated pressure, the simulated pressure levels (Fig. S9D) remain well below those that could damage the bacteria (*41-43*) or the microfluidic structures (*44, 45*), whereas bypass-loading operates under much gentler pressure conditions (Fig. S8D).

Beyond *E. coli*, the applicability of the cell-collection microfluidics to other UTI-related pathogens is demonstrated by the successful capture of *K. pneumoniae* and *S. aureus* in the microchambers (Fig. S10).

Cell enrichment in a small chamber significantly accelerated the pH change induced by *E. coli* metabolism. The pH response of cultivating a $5 \times 10^6$ cells/mL sample was measured under two conditions: (i) bulk cultivation in a 100 µL PDMS container (Detail setup is in Fig. S11) (*35*) and (ii) microfluidics cultivation in the LOSC chambers after 8 minutes of bypass-loading, resulting in an average of 8 cells per chamber. Both experiments were conducted in Luria-Bertani (LB) medium supplemented with 1 wt.% glucose at 37 °C. The pH signal, reflected by $\Delta V_{th}$, changes much more rapidly in the microfluidic cultivation than in bulk cultivation, as shown in Fig. 2C. This acceleration can be attributed to volumetric confinement: trapping on average ~8 cells in a 35 pL chamber corresponds to an effective volumetric density of approximately $3 \times 10^8$ cells/mL (Fig. 2B), which is nearly two orders of magnitude higher than the effective volumetric density used in the bulk cultivation (~$5 \times 10^6$ cells/mL).

**2.2 Reduction of test time with pH-unbuffered media**

LB media contains numerous amino acids and peptides, which may have pH buffering capacity. A stepwise acid titration test (Fig. S12) confirmed that LB retains a moderate pH buffering capacity, potentially slowing down the metabolism-induced pH changes. We therefore decided to investigate whether the response time could be shortened even further by replacing LB with an unbuffered medium. The experimental procedures followed those described in Section 4.5. Culturing *E. coli* in the microfluidic chip (~8 cells per chamber) using physiological 0.9% NaCl with 1% glucose resulted in a 60 mV change in 20 minutes, as compared to 50 min in LB media (Fig. 3A). This demonstrates that reducing the buffering capacity of the culture media can accelerate pH change.

We next investigated how fast we could observe antibiotic susceptibility using the NaCl media. To assess the susceptibility of *E. coli* (MG1655) to Ampicillin (AMP), the $\Delta V_{th}$ vs. time response was monitored for the samples cultured in LB and NaCl media (add 1 wt.% glucose) with and without AMP. In LB, both treated and untreated *E. coli* reached $\Delta V_{th}$ of 60 mV in 60 min, but whereas the untreated control sample continued to increase in $\Delta V_{th}$ to 120 min, the acidification halted in the AMP-treated sample after 70 minutes (Fig. 3B), suggesting that the antibiotic inhibited the metabolic activity of *E. coli*. Viable counts confirmed decreased *E. coli* viability in LB with AMP (Fig. S13A).

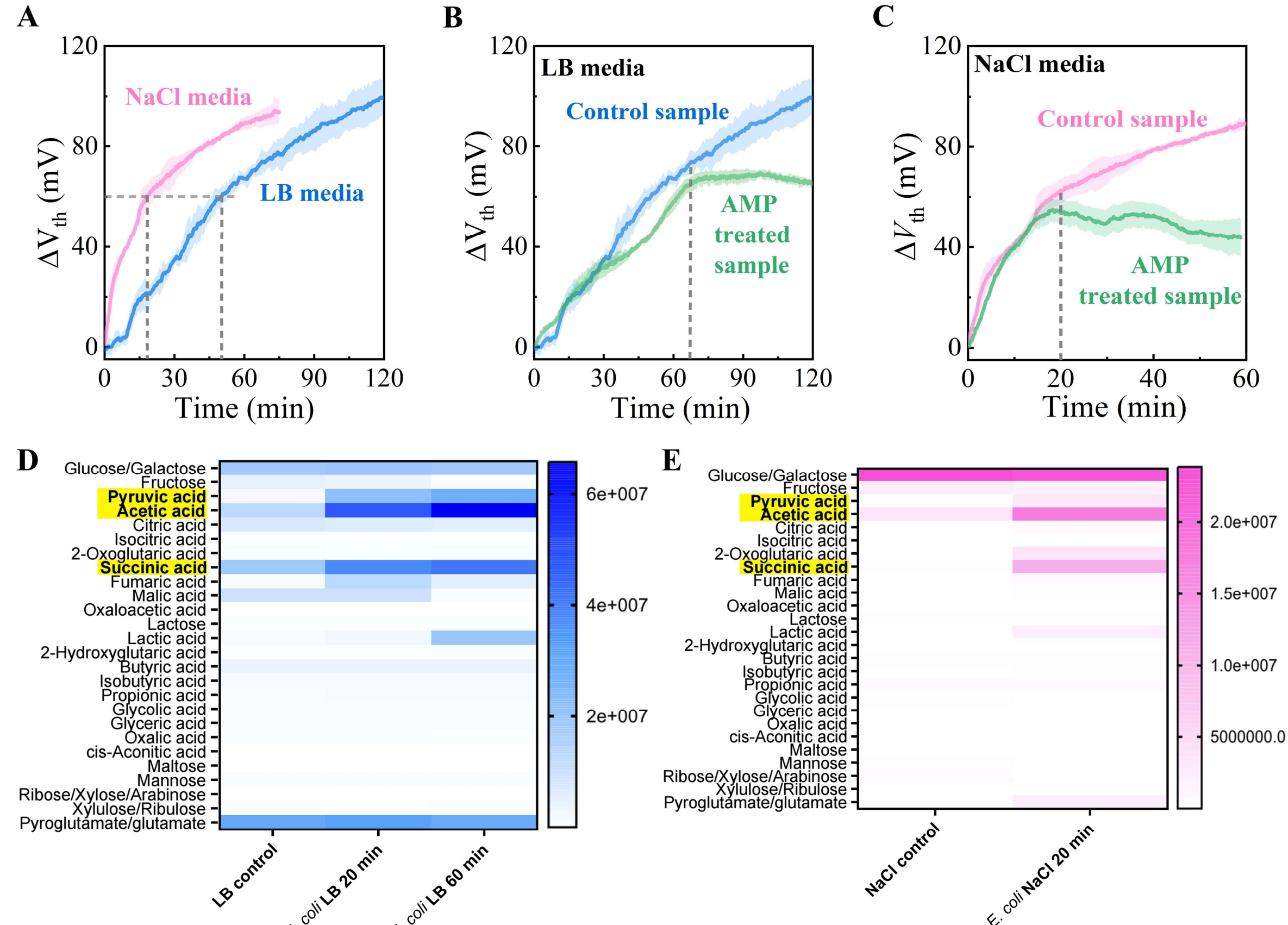

**Fig. 3 ASTs for *E. coli* (MG1655) using the LOSC platform with different test media.** (**A**) $\Delta V_{th}$ vs. time comparison for *E. coli* cultivation (~8 cells per chamber) in LB and NaCl media. (**B**) $\Delta V_{th}$ vs. time for *E. coli* in response to AMP (30 µg/mL) in LB media. (**C**) $\Delta V_{th}$ vs. time for *E. coli* in response to AMP (30 µg/mL) with NaCl media. The shaded areas represent the standard deviations calculated from three SiNWFET sensors. (**D**) Organic acid concentrations in LB media containing *E. coli* at $10^9$ cells/mL after 20 and 60 minutes of cultivation. (**E**) Organic acid concentrations in NaCl media containing *E. coli* at $10^9$ cells/mL after 20 minutes of cultivation.

In NaCl, both cultures reached $\Delta V_{th}$ of around 60 mV in 20 min, after which the $\Delta V_{th}$ continued to increase for the untreated *E. coli*, whereas acidification was halted for the AMP-treated *E. coli* (Fig. 3C). Viable counts also confirmed decreased viability of the treated culture in NaCl (Fig. S13B). The earlier halt in acidification observed in NaCl suggests that the use of unbuffered media can significantly accelerate AST.

To investigate whether the acidification of the media indeed arose from metabolic byproducts, we performed metabolomics analysis on the culture supernatant of *E. coli*. In both media, media acidification mainly originated from the secretion of acetate, pyruvate, and succinate (Fig. 3D and 3E) as expected for *E. coli* during overflow metabolism (*46*). Further, OD measurements and microscopy (Fig. S14) showed no apparent reduction in cell density over the two-hour culture, indicating that the pH changes arise from metabolism rather than cell lysis. Interestingly, less acids were produced in NaCl (Fig. 3E) as compared to LB at 20 min (Fig. 3D), even though a higher acidification was observed at 20 min (Fig. 3C). This further supports our findings that LB has a buffering capacity.

## 2.3 Rapid ASTs demonstration

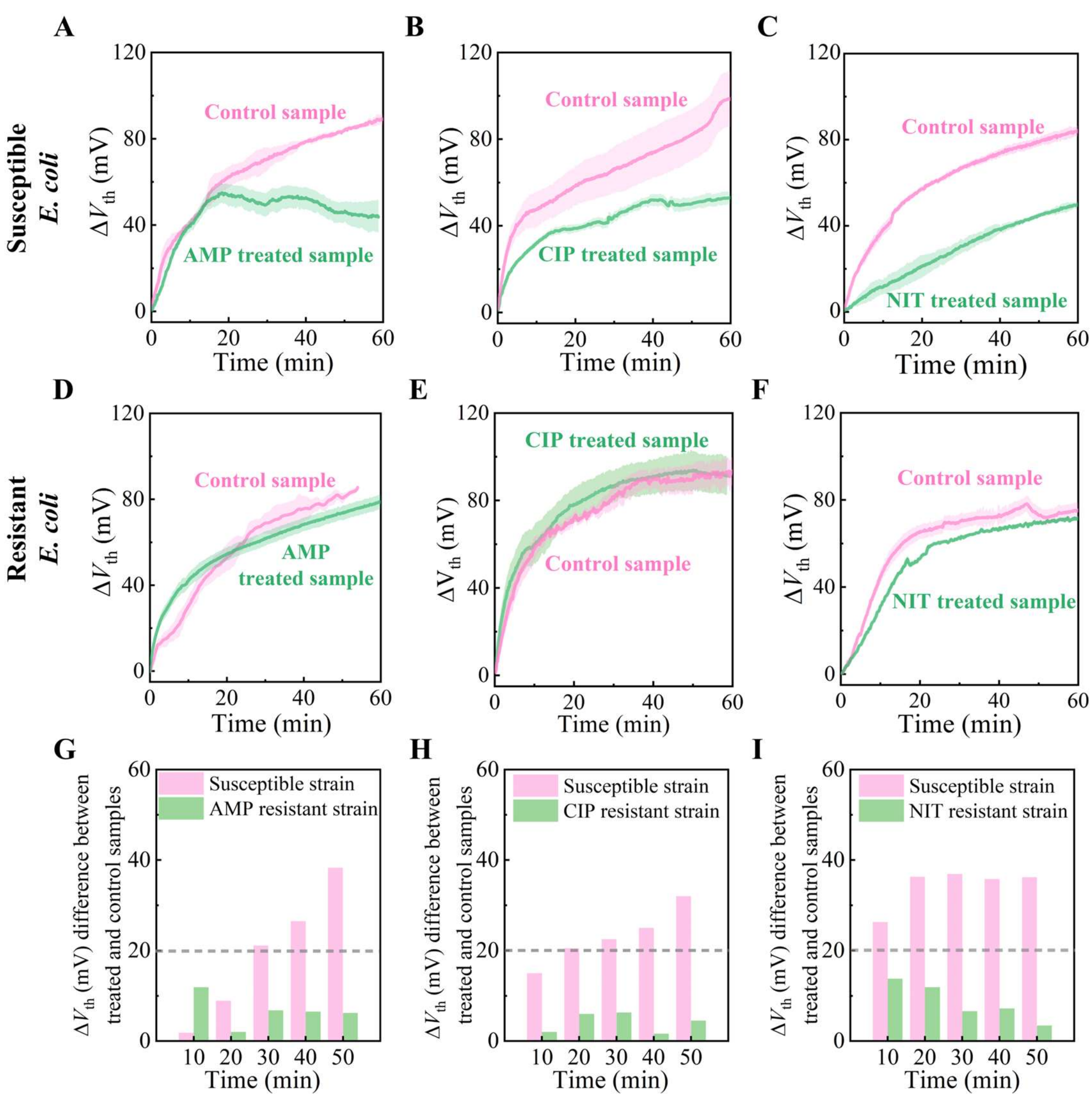

**Fig.4 ASTs of *E. coli* using the LOSC platform in 0.9 wt.% NaCl (1 wt.% glucose).** $\Delta V_{th}$ vs. time for susceptible (**A-C**) and resistant (**D-F**) *E. coli* in response to (**A** and **D**) AMP (30 μg/mL), (**B** and **E**) CIP (2 μg/mL), and (**C** and **F**) NIT (200 μg/mL). The shaded areas represent the standard deviations calculated from three SiNWFET sensors. (**G-I**) $\Delta V_{th}$ difference between treated and control samples over time for the susceptible and resistant strains under the same antibiotic conditions (calculated from **A-F**): (**G**) AMP (30 μg/mL), (**H**) CIP (2 μg/mL), and (**I**) NIT (200 μg/mL). The dashed lines indicate a 20 mV $\Delta V_{th}$ difference.

Next, we used the LOSC to demonstrate rapid ASTs of lab strains of *E. coli* to three clinically used antibiotics against urinary tract infections (UTIs): AMP, Ciprofloxacin (CIP), and Nitrofurantoin (NIT) in NaCl. The AST procedures followed those described in Section 4.5.

All three antibiotics reduced susceptible *E. coli* acid production, indicating suppressed metabolic activity. To define a quantitative susceptibility criterion, we used a $\Delta V_{th}$ difference threshold of 20 mV between treated and control samples, corresponding to the maximum standard error of $\Delta V_{th}$ for biological replicates in the LOSC AST measurements (Fig. S16). The onset of susceptibility was defined as the earliest time point at which the $\Delta V_{th}$ difference exceeded this threshold (*17, 28*). Using this criterion, susceptibility was detected at 30 min for AMP, 20 min for CIP, and 10 mins for NIT (Fig. 4A to 4C). To validate the platform's ability to differentiate between antibiotic-susceptible strains and resistant strains, the same set of ASTs was performed using *E. coli* strains, resistant to AMP (SK4582), CIP (SK4686), and NIT (SK7394) strains (Fig. 4D to 4F). The resistant strains did not reach the threshold difference of 20 mV in $\Delta V_{th}$ between antibiotic-treated and control samples over the test timeframe (Fig. 4G to 4I), indicating that no metabolic arrest occurred in these resistant strains in response to the antibiotics. Validation through plating for viable cells confirmed that the antibiotics effectively killed susceptible *E. coli*, while the resistant strains remained viable during the treatments under the same time period (Fig. S15).

To evaluate the alignment of our pH-based AST with standard MIC measurements, we performed dose–response experiments using three *E. coli* strains with CIP MICs of 0.012, 0.19, and 0.75 μg/mL determined by E-test (Fig. S17). Our SiNWFET pH sensors were used to test these samples at a cell density of ~$3 \times 10^8$ cells/mL. Increasing antibiotic concentration led to a reduction in metabolic activity (Fig. S18 A-C), forming clear dose–response relationships. MIC-equivalent values were extracted using $\Delta V_{th}$ difference threshold-based criteria. As a general metric, susceptibility is defined when the $\Delta V_{th}$ difference exceeds 20 mV during the measurement period, yielding MIC-equivalent values of 2, 4, and 8 μg/mL (Fig. S18D) and showing linear correlation with E-test MICs ($R^2 = 0.99$). Moreover, we can even limit the test duration to 5 minutes and define susceptibility as $\Delta V_{th} > 20$ mV within this period, to enable faster determination of MIC-equivalent values. This yields Fast-MIC-equivalent values of 8, 16, and 32 μg/mL (Fig. S18E), which also show a strong linear correlation with E-test MICs ($R^2 = 0.99$). These consistent linear relationships obtained with both criteria demonstrate that the pH-based AST can be aligned with standard MIC measurements.

To evaluate whether the LOSC platform is applicable beyond *E. coli*, we performed AST measurement using *Staphylococcus aureus* (*S. aureus*), a Gram-positive bacterium. As shown in Fig. S19A, the untreated *S. aureus* exhibited a rapid increase in $\Delta V_{th}$ followed by saturation (~100 mV) after 30 min, while NIT-treated cells showed a reduced acidification rate and a lower plateau (~70 mV), indicating antibiotic-induced metabolic inhibition. The kinetics and separation in the metabolic response of treated/control samples were consistent with those observed for *E. coli* (Fig 4C). The susceptibility of *S. aureus* to NIT (Fig. S19B) was further confirmed by the E-test, as shown in Fig. S19B. These results demonstrate that our LOSC system can be used for both Gram-negative (*E. coli*) and Gram-positive (*S. aureus*) bacteria.

**2.4 Rapid AST of clinical isolates**

To assess the clinical applicability of the LOSC platform, we evaluated the antibiotic susceptibility of a clinical isolate of uropathogenic *Escherichia coli* (UPEC 536) strain to two antibiotics: CIP (2 μg/mL) and NIT (200 μg/mL). The experimental procedures followed those described in Section 4.5. The total sample-handling time was 10 mins (cell loading: 8 mins;

culture media introduction: 1 min; and oil sealing: 1 min). Antibiotics reduced UPEC strain acid production as shown in Fig. 5A and 5B, indicating suppressed metabolic activity. In both cases, a $\Delta V_{th}$ difference greater than 20 mV between treated and control samples was reached as early as 10 mins (Fig. 5C and 5D). Notably, together with a 10-minute sample-handling time, the LOSC platform was able to resolve the susceptibility response with a sample-to-result time of within 20 minutes, demonstrating its potential for rapid phenotypic AST of clinical isolates.

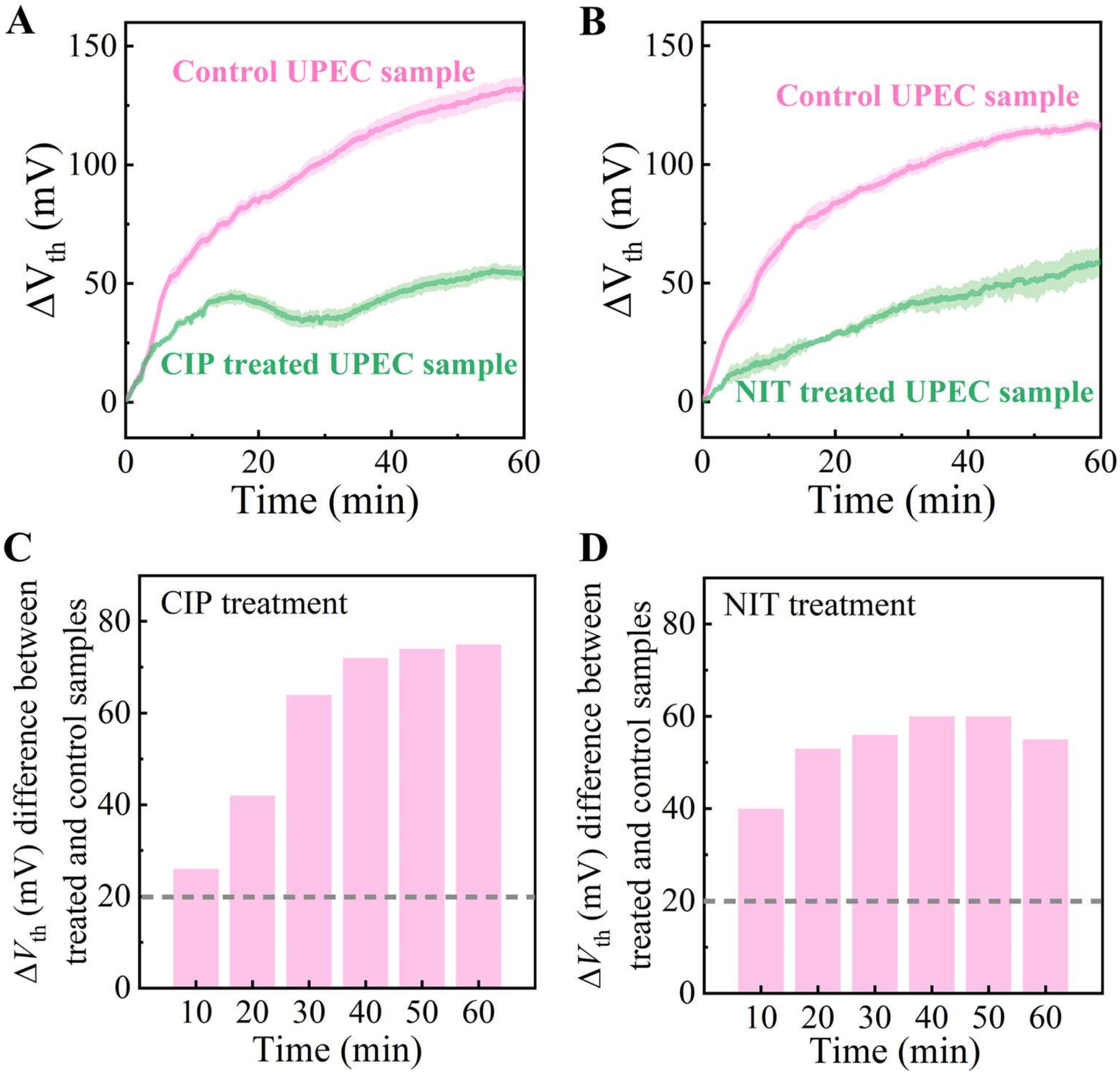

**Fig. 5 ASTs for UPEC using the LOSC platform in 0.9 wt.% NaCl (1 wt.% glucose).** $\Delta V_{th}$ vs. time for UPEC in response to (**A**) CIP (2 μg/mL) and (**B**) NIT (200 μg/mL). The shaded areas represent the standard deviations calculated from three SiNWFET sensors. $\Delta V_{th}$ difference between treated and control samples over time (calculated from A and B) in response to (**C**) CIP (2 μg/mL) and (**D**) NIT (200 μg/mL). The dashed lines indicate a 20 mV $\Delta V_{th}$ difference.

# 3. DISCUSSION

UTIs are the most frequent reason for antibiotic prescriptions. In standard clinical workflow (Fig. S20), AST requires multiple steps of cultivation (*11*), making the total sample-to-result time somewhere between one to three days (*47, 48*). This is too long to guide treatment choices. Instead, the prevalence of resistance is used to determine which antibiotic to currently use. A prevalence higher than 25% of resistant isolates can result in changed guidelines of treatment (*49*). This results in an overuse of last-resort antibiotics, when in most cases a cheaper, older antibiotic would be fully functional. Rapid ASTs could reduce the use of last-resort antibiotics, to a reduced cost for both the patient and society.

The long sample-to-result time mainly originates from the pre-culturing step required by conventional ASTs. Table 1 compares representative AST platforms in terms of working principle, sample compatibility, and testing time. Commercial clinical systems such as VITEK 2, Phoenix, and Sensititre are extensively validated and robust, but they require pure isolates and pre-cultivation, adding 24-48 hours to the total sample-to-result times. Many recently reported rapid AST approaches can shorten the testing time, but most still require a pre-culture step, which dominates the overall workflow and keeps the true sample-to-result time in the range of several hours to one or two days. A few AST platforms listed in Table 1 can perform direct-from-sample tests by integrating microfluidic(*17, 28, 39*) or magnetic-bead-based bacterial enrichment (*27*), and significantly shorten the sample-to-result time. Most of these systems rely on optical readouts based on bacterial growth, whereas our LOSC platform employs a purely electrical readout to profile cellular metabolism. Since our approach does not rely on observable morphological changes, it has the potential to be faster than optical methods that monitor cell growth. It also provides a different phenotypic dimension and offers advantages in integrability and automation.

Against this background, the LOSC platform holds great promise for rapid AST directly on urine samples without the need for any preculture. To ensure efficient bacteria loading from urine, we also incorporated a filter region (*17*) at the inlet of the LOSC to prevent, e.g., white blood cells and debris from entering the microfluidic channels (Fig. S3). Clinical UTIs are diagnosed from $10^3$ to $10^5$ cells/mL, depending on country (*50, 51*). For UTI samples with > $10^4$ cells/mL, which is in the lower range for clinically relevant UTIs, sufficient bacteria loading (>3 cells per chamber) can be achieved within 10 minutes. Thus, our LOSC platform could, within 10 minutes, determine whether there is a clinically relevant infection at all. For severe bacterial infections, with more than $10^5$ cells/mL, loading could be even faster. For samples containing fewer than $10^4$ cells/mL, the bacterial loading step may become the time-limiting step in the LOSC workflow. To address this, loading efficiency could be further improved by increasing the flow rate or reducing the volume of the culture chambers. Or, the samples could be centrifuged to concentrate the sample prior to loading.

Moreover, the modular architecture of the LOSC platform permits spatial tiling of microfluidic sensor units, supporting future expansion toward high-throughput configurations. The electrical readout is naturally compatible with multiplexed architectures, supporting future scaling toward higher-throughput implementations.

While this study primarily focused on AST for *E. coli*, the applicability of the LOSC platform to a Gram-positive bacterium, *S. aureus*, was also demonstrated (Fig. S19). The LOSC platform can be further extended to other microbial species. Specifically, for slow-growing bacteria such as *Staphylococcus aureus* small colony variants (SCVs) (*52*) and *Mycobacterium tuberculosis* (*53, 54*), our metabolism-based AST, which monitors acid production rather than growth rate, has the potential to deliver results significantly faster than growth-based rapid AST methods. And as metabolic profiles of different pathogens vary (35), the LOSC-generated data could in the future be used for infection diagnosis and AST results concurrently in the same run.

Another promising direction for future development is the design of microfluidics capable of isolating individual bacterial cells for single-cell AST. High-throughput single-cell AST would enable detailed characterization of heterogeneity in antibiotic susceptibility within bacterial populations (*55-57*) and offer new insights into the link between metabolic state and resistance (*34, 58-60*). Moreover, in polymicrobial infections, single-cell analysis could help resolve species-specific susceptibility profiles, ultimately guiding more precise and effective antimicrobial therapies.

In summary, the LOSC AST platform offers an electrical, low-cost solution capable of performing rapid AST directly on UTI samples, achieving a sample-to-result time of under 20 minutes. This makes it a promising candidate for point-of-care diagnostics and timely clinical decision-making.

**Table 1. Comparison with representative AST platforms in terms of sensing principle, sample compatibility, testing time, and technology readiness level (TRL).**

| Principle | Sample compatibility | Testing time | TRL | Commercial name/ Ref. |
|---|---|---|---|---|
| Light absorbance measurement | Pure culture | 5-18 h | 9 | VITEK 2 (*61*) |
| Light absorbance measurement | Pure culture | 4-16 h | 9 | Phoenix$^{TM}$ (*62*) |
| Fluorescence detection | Pure culture | 20-24 h | 9 | Sensititre$^{TM}$ (*63*) |
| Bacterial nanomotion analysis | Short-term culture (6 h) | 2 h | 9 | Resistell$^{TM}$ (*64*) |
| Bacterial mass detection | Positive culture | 4.5 h | 9 | LifeScale (*65*) |
| Single bacterial replication image | Urine specimen | 1 h | 9 | PA-100 AST (*66*) |
| Droplet shrinkage (subpopulations detection) | Pure culture | 12 h | 4 | (*57*) |
| Magnetic beads enrichment and microculture | Blood specimen | 4-12 h | 6 | (*27*) |
| Raman-assisted detection | Urine specimen | 3 h | 6 | (*23*) |
| SiNWFET-based pH monitoring | Pure culture | 0.5 h | 4 | (*35*) |
| **This work** (microfluidic enrichment and pH monitoring) | Urine specimen | 10 min | 4 | |

**Testing time** refers to the duration of the AST test itself. Methods requiring pure culture samples typically involve 24–48 h pre-cultivation of clinical samples. **Technology readiness level (TRL)** indicates technology maturity on a scale from 1 (early-stage concept) to 9 (fully validated and commercialized system) (*7*).

# 4. MATERIALS AND METHODS

## 4.1 Silicon sensor chip realization

The SiNWFET sensors were fabricated as previously reported (*35, 67*), using standard silicon process technology on SOI wafers with a 55 nm thick, p-type ($1\times10^{15}$ $cm^{-3}$) doped top Si layer. The process flow is illustrated in Fig. S1. In brief, the wafer was heavily n-doped by arsenic (As) implantation while the channel region was protected by photoresist during the implantation. The silicon nanowires with source and drain pads were formed by electron beam lithography (EBL) and reactive ion etching (RIE). To reduce series resistance, NiSi was formed on the $n^+$-Si leads connecting the SiNWFETs to the probe pads at the edges of the chip. Subsequently, a 4-nm-thick $HfO_2$, deposited by atomic layer deposition (ALD), served as both gate dielectric and liquid passivation. Finally, forming gas annealing (5% $H_2$/95% $N_2$) at 400 °C for 30 minutes passivated oxide/silicon interface traps.

For AgCl/Ag p-REs integration (*40*), a 150 nm Ag layer with a 10 nm Ti adhesion layer was deposited via evaporation and patterned by lift-off. Photoresist (UVN 2300-0.5) covered the electrodes, with contact holes opened by EBL to expose Ag near the SiNWFETs, as shown in Fig. 1A. The exposed Ag was further treated with a $FeCl_3$ solution (0.1M) for 30 seconds to form an AgCl layer.

## 4.2 Cell-collection microfluidics fabrication

The cell-collection microfluidics were made from polydimethylsiloxane (PDMS) (Silgard 184; Dow Corning) using a silicon mold fabricated using lithography and RIE. Each PDMS chip contains two independent microfluidic testing units, and each unit comprises a culture-chamber array consisting of 27 chambers. Details of the fabrication process are shown in Fig. S2 and S3.

## 4.3 Bonding of silicon sensor chip with cell-collection microfluidics

We devised an aligner to align a SiNWFET and its pairing AgCl/Ag p-REs within each microchamber in the cell-collection microfluidics, as shown in Fig. 1B. Before alignment, both the silicon sensor chip and the cell-collection microfluidics underwent $O_2$ plasma treatment for surface activation. The details of the alignment and bonding process are represented in Fig. S4.

## 4.4 Bacterial strains

The susceptible *E. coli* strain used in this study was *Escherichia coli* K12 MG1655. The UPEC strain used in this study was UPEC536. The genotypes of each strain are listed in the table. S2.

## 4.5 AST protocol

During AST, *E. coli* samples at an initial concentration of $1 \times 10^6$ cells/mL were introduced into the LOSC device via a syringe pump operating at a flow rate of 8 μL/min for 8 minutes, using the bypass loading mode. Under these loading conditions, the microchambers were expected to reach an average volumetric cell density of approximately $3\times10^8$ cells/mL (see Fig. 2B). Following cell loading, test media, with or without antibiotics, were perfused into the LOSC at the same flow rate (8 μL/min) for 1 minute. Subsequently, FC-70 oil was introduced into both the front and back channels at a flow rate of 0.5 μL/min for 1 minute to effectively seal the chambers. Throughout the AST process, the LOSC is placed on a chip holder, which is maintained at a constant temperature of 37 °C. The surface of LOSC is covered by FC-70 oil to further prevent liquid evaporation.

### 4.6 Electrical measurement

Electrical measurements were carried out using an HP4155A semiconductor parameter analyzer. HP4155A is connected to the LOSC via a probe card for drain current measurement. Multiple SiNWFETs were measured simultaneously via a Keysight 34970A switch unit. First, the transfer characteristics $I_D$-$V_G$ were recorded. Subsequently, the drain current $I_D$ was monitored in real time under a constant source-drain voltage $V_{DS}$ of 0.3 V and a constant gate voltage $V_G$ applied relative to the on-chip AgCl/Ag p-RE. $V_{DS}$ was set to 0.3 V to ensure that the SiNWFET operates in the saturation regime, thereby reducing $I_D$ sensitivity to $V_{DS}$ fluctuations and improving readout stability. During measurement, $V_G$ was set within the subthreshold region. The recorded $I_D$ values were then converted to $\Delta V_{th}$ using the previously acquired transfer curve (*40*).

### 4.7 *E. coli* metabolic byproducts analysis

Screening of 3-nitrophenylhydrazine (3-NPH) derivatized metabolites by liquid chromatography coupled to mass spectrometry (LC-MS) was performed at the Swedish Metabolomics Center in Umeå, Sweden. Information about reagents, solvents, standards, reference and tuning standards, and stable isotopes internal standards can be found in the Supplementary Materials.

## Supplementary Materials

The PDF file includes:
Supplemental Texts
Figs. S1 to S20
Tables S1 to S2

**Acknowledgments:** The device fabrication was done in the Ångström Microstructure Laboratory (MSL) at Uppsala University, and the technical staff of MSL are acknowledged for their process support. Swedish Metabolomics Centre, Umeå, Sweden (www.swedishmetabolomicscentre.se) is acknowledged for the screening of 3-NPH derivatized metabolites by LC-MS. The authors acknowledge the use of ChatGPT-5.5 (developed by OpenAI) to assist with language editing and refinement during the preparation of this manuscript. **Funding:** This work was partially supported by Olle Engkvist Foundation 214-0322, the Swedish Research Council under Grant VR 2019-04690, 2023-02925_3, and 2024-05176, and the Wallenberg Academy Fellow program under Grant KAW 2020-0190. **Author Contributions:** Conceptualization: Z.X., S.K., and Z.Z. Methodology: Z.X., Y.Y, S.K., and Z.Z. Investigation: Z.X., Y.Y, V.N., and Z.Z. Data curation: Z.X. and V.N. Validation: Z.X. and Z.Z. Writing-original draft: Z.X. Writing-review & editing: Z.X, V.N., Y.Y., S.K., and Z.Z. Formal analysis: Z.X., S.K., and Z.Z. Visualization: Z.X. Software: Y.Y. Resources: Z.Z. P.M. and S.K. Funding acquisition: S.K. and Z.Z. Project administration: S.K. and Z.Z. Supervision: S.K. and Z.Z. **Competing interests**: The authors declare that a patent application PCT/EP2024/051148 (WO 2024/156586 A1) related to this work has been filed by Zhen Zhang. The patent application is currently pending. **Data, code, and materials availability**: All data and code needed to evaluate and reproduce the results in the paper are present in the paper and/or the Supplementary Materials. This study did not generate new materials

Supplementary Materials for

# A lab-on-a-silicon-chip platform for all-electrical antibiotic susceptibility tests with a sample-to-results time within 20 minutes

Zheqiang Xu *et al.*

*Corresponding author: Sanna Koskiniemi, sanna.koskiniemi@icm.uu.se; Zhen Zhang, zhen.zhang@angstrom.uu.se

**This PDF file includes:**

# Supplementary Text

## 1. Silicon sensor chip fabrication

The SiNWFETs were fabricated on a SOI wafer using standard silicon process technology. A schematic of the fabrication process flow is depicted in Fig. S1. The key process steps are summarized below:

I: The initial lightly p-type doped ($1\times10^{15}$ $cm^{-3}$) SOI layer was 55 nm thick, on top of a 145 nm thick buried $SiO_2$ layer.

II: The wafer was heavily n-doped by arsenic (As) implantation (energy = 7 keV, dose = $5\times10^{15}$ $cm^{-2}$) while the channel region was protected by photoresist during the implantation. The As dopants were activated by rapid thermal processing (RTP) in $N_2$ at 1000 °C for 10 s.

III: The device structure, including source (S) / drain (D) pads, connecting lead, and nanowire channel (1×0.1 μm), was defined using electron-beam lithography (EBL) and subsequent reactive-ion etching (RIE).

IV: To further diminish the series resistance, a 5-nm-thick Ni layer was deposited on the S / D regions and connecting lead via a lift-off process and subsequently annealed at 400 °C in $N_2$ for 30 s using a RTP to form NiSi. Afterward, a 4-nm-thick $HfO_2$ layer, grown by atomic layer deposition (ALD, R200 unit, Picosun) at 170 °C, served as the gate dielectric and the passivation layer for liquid operation. The probe pads along the chip edges were metalized with a bilayer of 10 nm Ti and 100 nm Al. Finally, forming gas annealing in a gas mixture of 5% $H_2$ and 95% $N_2$ at 400 °C for 30 min was used to passivate the oxide/silicon interface traps.

V: For the fabrication of the AgCl/Ag pseudo reference electrode (AgCl/Ag p-RE), a bilayer of 10 nm Ti and 150 nm Ag was deposited on the chip surface by a lift-off process. The Ti layer is designed to improve the adhesion of Ag to the $HfO_2$ layer. Subsequently, a multihole (with a hole diameter of 5 μm) photoresist protection layer was deposited above the Ag and hard-baked at 110 °C for 2 h. Finally, the p-RE was treated with a $FeCl_3$ (0.1 M) solution for 30 seconds to form a thin AgCl layer.

## 2. Cell-collection microfluidics fabrication

The cell-collection microfluidics were fabricated using PDMS molding, as illustrated in Fig. S2.

I: A silicon wafer served as the mold, with fine structures (exit and loading channels of culture chamber) formed using a combination of EBL and RIE. These structures have a depth of 1.5 μm. The loading channels are 2 μm wide, matching the typical size range of common UTI pathogens (*28, 50*) and thereby allowing their entry into the chambers. In contrast, the exit channels have a minimum width of 0.5 μm, serving as a size-based restriction that prevents most bacteria (*68-71*) from passing into the back channel while still permitting fluid flow.

II: Front and back channels, culture chamber bodies, and ports were lithographed onto a 20-μm-thick SU8 layer, with the microfluidic layout details shown in Fig. S3 (A and B). The width of the front and back channels is 100 μm. The dimension of the culture chamber is 35 × 50 × 20 μm (width × length × height), holding only 35 pL.

III: A mixture of PDMS base and curing agent (Sylgard 184, Dow Corning) at a 1:8 ratio was degassed under vacuum for 30 minutes. The Si mold surface was modified with a monolayer of trichloro(1,1,2,2-perfluorocyclyl) silane to prevent PDMS adhesion. After pouring the mixture into the mold, it was further degassed and cured at 60 °C for five hours.

IV: De-molding was followed by dicing the PDMS into individual PDMS chips and punching fluidic connection ports. The final pattern of PDMS is shown in Fig. S3D. Each PDMS chip contains two independent microchamber arrays, and each array consists of 27 chambers.

## 3. The bonding alignment system

To achieve robust integration between the silicon sensor chip and the cell-collection microfluidic layer, we employed an oxygen plasma bonding technique. An alignment platform (see Fig. S4) is designed to achieve fine alignment between the silicon sensor chip and cell-collection microfluidics during the bonding process.

The cell-collection microfluidics, made of PDMS, was first placed on a glass carrier and gently pressed to establish temporary adhesion, ensuring a flat surface for subsequent alignment. Both the PDMS microfluidics and the silicon chip were then treated with oxygen plasma to activate their bonding surfaces. It is worth noting that the bonding between metal and PDMS is weak. Covering the Ag layer with a photoresist protection layer significantly enhances bonding strength, ensuring leakless sealing.

For alignment, the sensor chip was mounted on the chip carrier positioned at the centre of an X/Y/Z/R alignment stage. The glass carrier holding the PDMS layer was fixed on a holding arm, with the activated bonding surface of the microfluidics facing downward. The Z-stage was raised until the distance between the two components was reduced to approximately 2 mm, allowing clear visualization of both surfaces under a microscope. Fine adjustments in the X, Y, and rotational axes were made to align the culture chambers of microfluidics precisely with the corresponding sensing areas on the silicon chip.

Once alignment was achieved, the Z-stage was further raised to bring the two surfaces into contact. After gentle bonding, the Z-stage was lowered to release the assembled chip. Then the assembled LOSC chip was placed on a 100 °C hotplate and annealed for 6 hours to ensure a strong bond.

## 4. SiNWFET noise measurement

The power spectral density (PSD) of the $V_G$-referred noise ($S_{VG}$) was characterized at different drain currents (10, 100, and 400 nA) with a constant $V_D$ of 0.3 V using a low-frequency noise analyzer (Keysight E4727A) coupled with a B1500A, over a frequency range of 0.1 Hz–1 MHz. The noise level showed no significant variation with drain current (Fig. S5). The root-mean-square (RMS) gate voltage noise was obtained by integrating the PSD, $RMS = \sqrt{\int_{0.1H}^{500Hz} S_{VG}(f)df}$. and was found to be < 1 mV, which is far below the metabolic signal observed in bacterial experiments (Fig. 4), indicating that device noise does not limit sensing performance.

## 5. *E. coli* distribution in each culture chamber

One-way ANOVA analyses were performed to evaluate the distribution of *E. coli* across culture chambers in the cell-collection microfluidics under both bypass and force loading modes, using *E. coli* samples with varying cell concentrations. In each loading experiment, the number of *E. coli* cells was manually counted in each chamber. The 24 chambers in each experiment were divided into three groups of eight chambers each based on their sequential positions along the flow path. For each group, the mean and variance of cell counts were calculated, and the statistical difference was determined using one-way ANOVA. The ANOVA results are shown in the Table. S1.

## 6. The numerical simulation of the flow profile in microfluidic

Numerical simulation methods were used to analyze the flow and pressure profile in the microfluidics during bacteria loading. The model used in the numerical simulation is shown in Fig. S7. The model's dimensions are consistent with the cell-collection microfluidics described in Fig.S3. The width of the front channel and the back channel is 100 μm, and the thickness of the channel is 20 μm. The culture chamber is 35×50×20 μm (width × length × thickness). The thickness of the loading channel and exit channel is 1.5 μm.

The equations that govern the physics of this flow field simulation are the incompressible, steady-state Navier-Stokes momentum equations. The equations were solved numerically by employing the finite element method using COMSOL Multiphysics software. During the simulation, we used the parameters for water from the COMSOL library. For the fluid flow, the wall's no-slip boundary condition was applied. For the inlet of the model (shown in Fig.S8(A), the boundary condition to fully developed flow with an average velocity of 0.066 m/s (corresponding to the flow rate of 8 μL/min) was applied. For the outlets of the model, the boundary condition is pressure equal to 0.

## 7. Bulk cultivation setup

To evaluate the acceleration effect of the bacteria enrichment on metabolism-induced pH changes, a bulk cultivation setup was used as a control (Fig. S11). All conditions (temperature, sample, sensor, and readout method) were identical to the microfluidic experiments; the only difference was the absence of microfluidic confinement.

For a bulk cultivation setup, a 100 μL PDMS container was placed on an SiNWFET chip with an on-chip AgCl/Ag p-RE. The chip was placed on a temperature-controlled holder and maintained at 37 °C. Before the experiment, 50 μL of LB medium was added for calibration and then removed. It was replaced with 100 µL of an *E. coli* suspension in LB at $5 \times 10^6$ cells/mL. The container was sealed with 20 µL of FC-70 oil to minimize evaporation (Fig. S11). During the experiment, the drain current was continuously recorded and converted into threshold voltage-$V_{th}$ shift using the pre-measured I–V characteristics.

## 8. pH buffering capacity of the culture media

To investigate the buffering capacity of LB, a PDMS container was filled with 100 µL of pH 7 LB solution, and the pH changes were monitored using the SiNWFET sensor (Fig. S11). During the experiment, 10 µL of the solution was periodically removed and replaced with 10 µL of 0.1 M KCl solution with a known pH. This method was intended to gradually shift the pH of the container's solution. For example, adding a pH 4 KCl solution should theoretically lower the solution pH to 5, in the absence of pH buffering capacity.

## 9. *E. coli* time-kill under antibiotic treatments in different media

*E. coli* cultures were grown overnight in LB at 37 °C with shaking at 200 rpm. On the day of testing, 100 µL of overnight culture was centrifuged at 1500 × g (rcf) for 4 mins and resuspended in 1.5 mL of LB with 1% glucose or 0.9% NaCl with 1% glucose. Antibiotics were added to make the final concentration required for testing. For each time point, 20 µL was removed and serially diluted in 180 µL PBS, and 10 µL was spotted onto LB agar. The LB agar plates were cultured overnight at 37 °C, and colonies were counted. The survival percentage was calculated by dividing the different time points by the initial count and multiplying by 100. All data were processed with Graphpad Prism 10.

## 10. *E. coli* metabolic byproducts analysis

10.1 Sample preparation and 3-NPH derivatization:

The samples were derivatized according to a published method (*72*) with some modifications. 20µl of bacteria media was transferred to an liquid chromatography (LC) plate (96 Agilent WP 5043-9314 0.33 mL VBtm) and spiked with 5 µl of internal standard mix (including 13C9-Phenylalanine, 13C3-Caffeine, D4-Cholic acid, 13C9-Caffeic Acid, D6-Salicylic acid, D4-Succinic acid, 13C5,15N-L-glutamic acid, 13C6-D-glucose, 13C5-L-proline, 13C4-alpha-ketoglutarate, 13C4-Malic acid, 13C4-Fumaric acid, 13C4-2-oxoglutaric acid, D4-Citric acid, 13C3-Puryvic acid, 13C3-Lactic acid, 13C5-2-hydroxyglutaric acid and 13C5-Itaconic acid). 20 µL of 120 mM EDC (dissolved in 6% pyridine in 50% acetonitrile) and 20 µL of 200 mM 3-NPH (dissolved in 50% acetonitrile) were consecutively added to the samples. The plate was incubated at room temperature (21 °C) for 60 minutes and followed by centrifugation at +4 °C, 4000 rpm (3488 g), for 10 minutes on a tabletop swingout rotor. Afterwards, 40 µL of the supernatant was transferred to a new LC plate, and 60 µl 0.05 mg/mL BHT (dissolved in pure methanol) was added to the samples. A small aliquot of the bacterial media was pooled and used to create quality control (QC) samples. MS/MS analysis was run on the QC samples for identification purposes. The samples were analyzed according to a randomized run order.

10.2 LC-MS analysis:

The chromatographic separation was performed on an Agilent 1290 Infinity UHPLC system (Agilent Technologies, Waldbronn, Germany). 1 µL of each sample was analyzed on a Luna Omega Polar C18, 100 × 2.1 mm, 1.6 µm in combination with a 2.1 mm SecurityGuard cartridge (Phenomenex, Torrance, CA, USA) by using gradient elution of 0.1% formic acid (*v*/*v*) in water as mobile phase A and acetonitrile/isopropanol (70/30, *v*/*v*) as mobile phase B. The mobile phase was delivered on the column by a flow rate of 0.4 mL/min with the following gradient: 0-1 min (5% B), 5 min (30% B), 9 min (50% B), 12 min (78% B), 15 min (95% B), 16 min (100% B), 18 min (100% B), 18.1 min (5% B), 20 min (5% B). The column and autosampler were thermostated at 40 °C and 4 °C, respectively.

The compounds were detected with an Agilent 6546 Q-TOF mass spectrometer equipped with a jet stream electrospray ion source operating in positive or negative ion mode. The settings were kept identical between the modes, with the exception of the capillary voltage. A reference interface was connected for accurate mass measurements; the reference ions purine (4 µM) and HP-0921 (Hexakis(1H, 1H, 3H-tetrafluoropropoxy)phosphazine) (1 µM) were infused directly into the MS at a flow rate of 0.05 mL/min for internal calibration, and the monitored ions were purine m/z 121.05 and m/z 119.03632; HP-0921 m/z 922.0098 and m/z 966.000725 for positive and negative mode respectively. The gas temperature was set to 150°C, the drying gas flow to 8 L/min, and the nebulizer pressure to 35 psi. The sheath gas temperature was set to 350 °C, and the sheath gas flow was 11 L/min. The capillary voltage was set to 4000 V in positive ion mode and to 4000 V in negative ion mode. The nozzle voltage was 300 V. The fragmentor voltage was 120 V, the skimmer 65 V, and the OCT 1 RF Vpp 750 V. The collision energy was set to 0 V. The m/z range was 50 - 1000, and data were collected in centroid mode with an acquisition rate of 4 scans/s (2265 transients/spectrum).

10.3 Data analysis-evaluation/statistical methods:

The data processing was performed using Agilent Masshunter Profinder version B.10.0.2 (Agilent Technologies Inc., Santa Clara, CA, USA). An in-house LC-MS library of 3-NPH derivatized metabolites built up by authentic standards run on the same system with the same chromatographic and mass-spec settings was searched for using the Batch Targeted feature extraction with an m/z tolerance of 20 ppm and a retention time window of 0.1 min. The identification of the metabolites was based on exact mass and retention time match.

## Supplementary figures

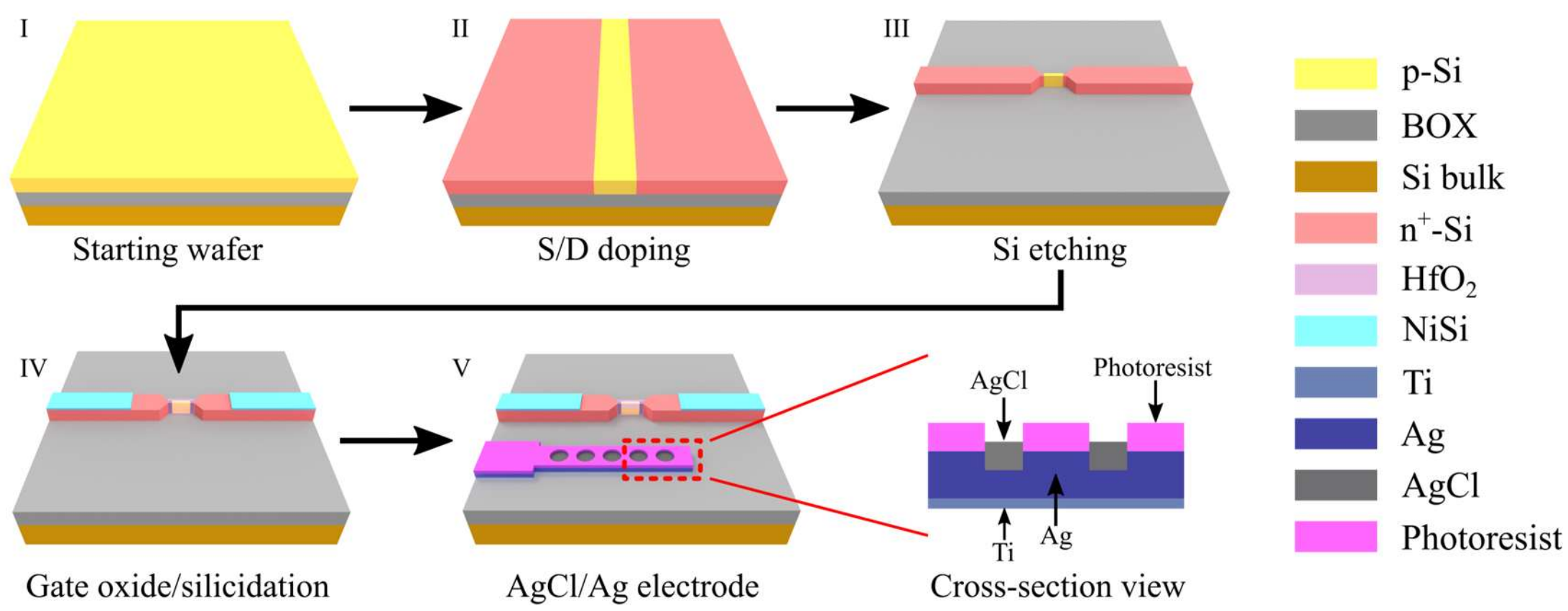

**Fig. S1 Fabrication process flow of the SiNWFET sensor.** (I) Starting with a 4-inch SOI wafer. (II) Source/drain region formation via ion implantation. (III) Definition of Si nanowire structure through RIE. (IV) Silicidation of source/drain and interconnects, followed by gate oxide deposition. (V) Fabrication of the on-chip AgCl/Ag p-RE.

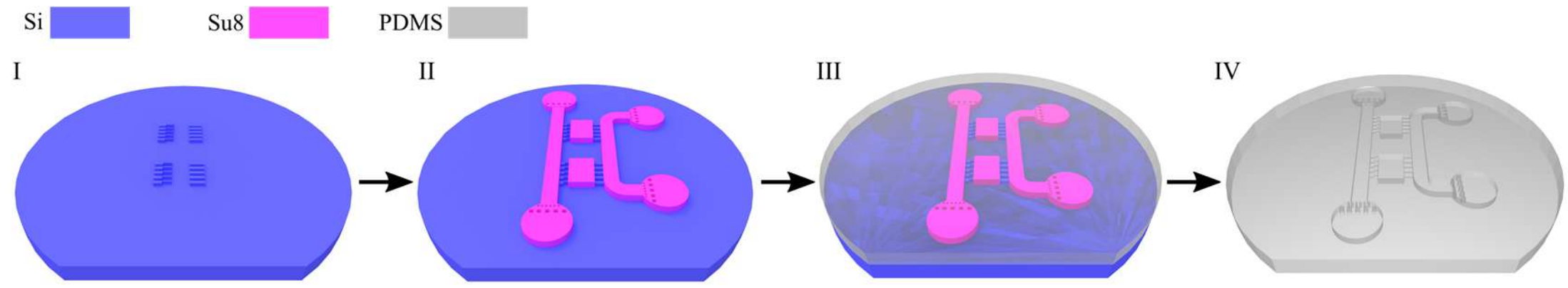

**Fig. S2. Fabrication process flow for cell-collection microfluidics.** (I) Fabrication of fine structures on a Si mold using EBL and RIE. (II) Formation of large structures on the same Si mold through SU8 photolithography. (III) Pattern transfer from the Si mold to the PDMS substrate. (IV) The final PDMS-based microfluidics.

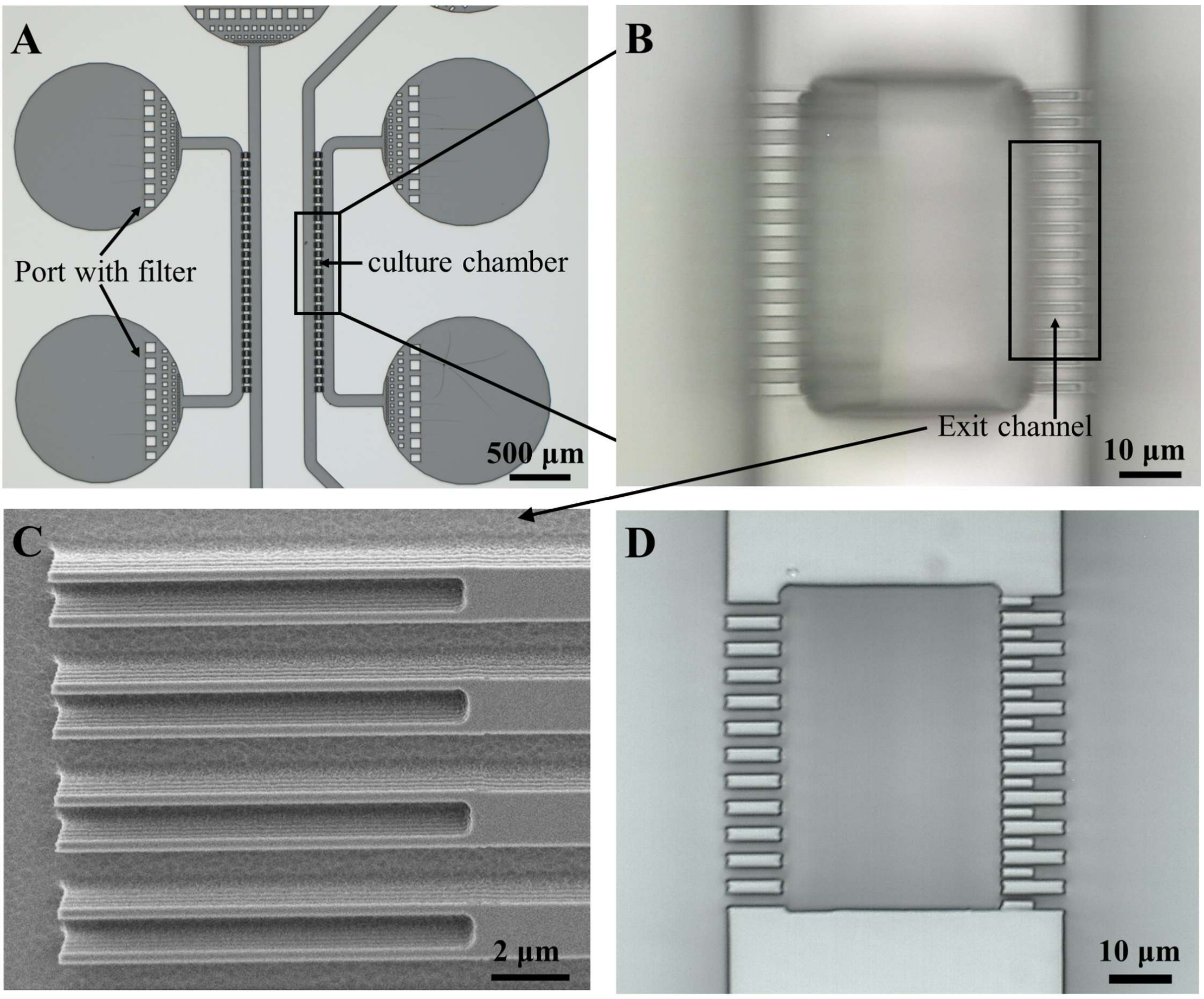

**Fig. S3 PDMS molding process.** (**A**) An optical image of the Si mold for the cell-collection microfluidics fabrication. Each cell-collection microfluidics contains two columns of culture chambers. So, it can run two parallel experiments at the same time. Each array consists of 27 chambers. To prevent the flow of large particles and capture large air bubbles, each fluidic port contains a filter region (*17*). (**B**) A zoomed-in optical image of a culture chamber on a Si mold. (**C**) A SEM image of the exit channel. The depth of the exit channel is 1.5 µm. (**D**) An optical image of a culture chamber fabricated on PDMS.

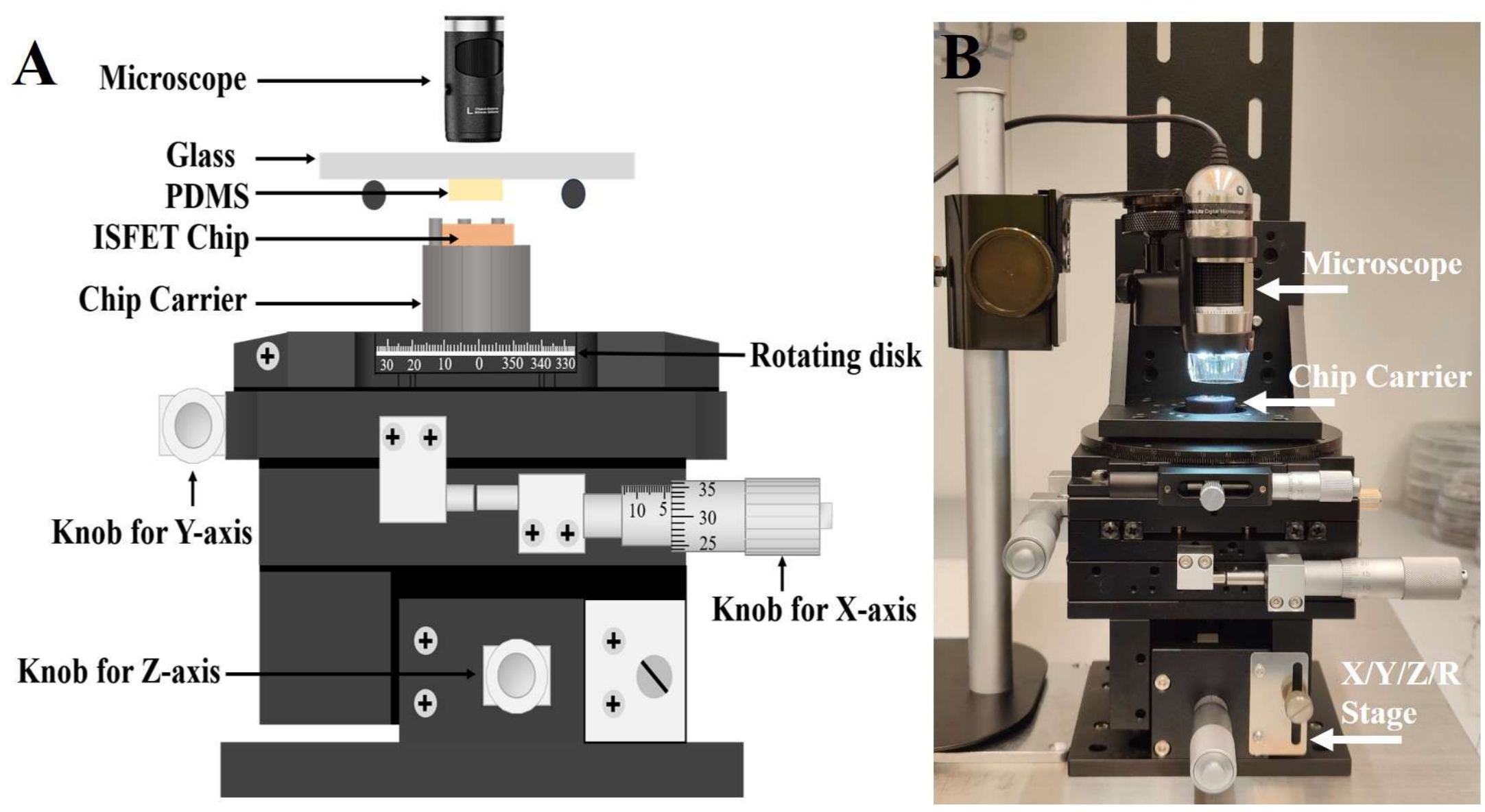

**Fig. S4 The bonding alignment system.** (**A**) Schematic of the bonding alignment system. (**B**) The image of the bonding alignment system.

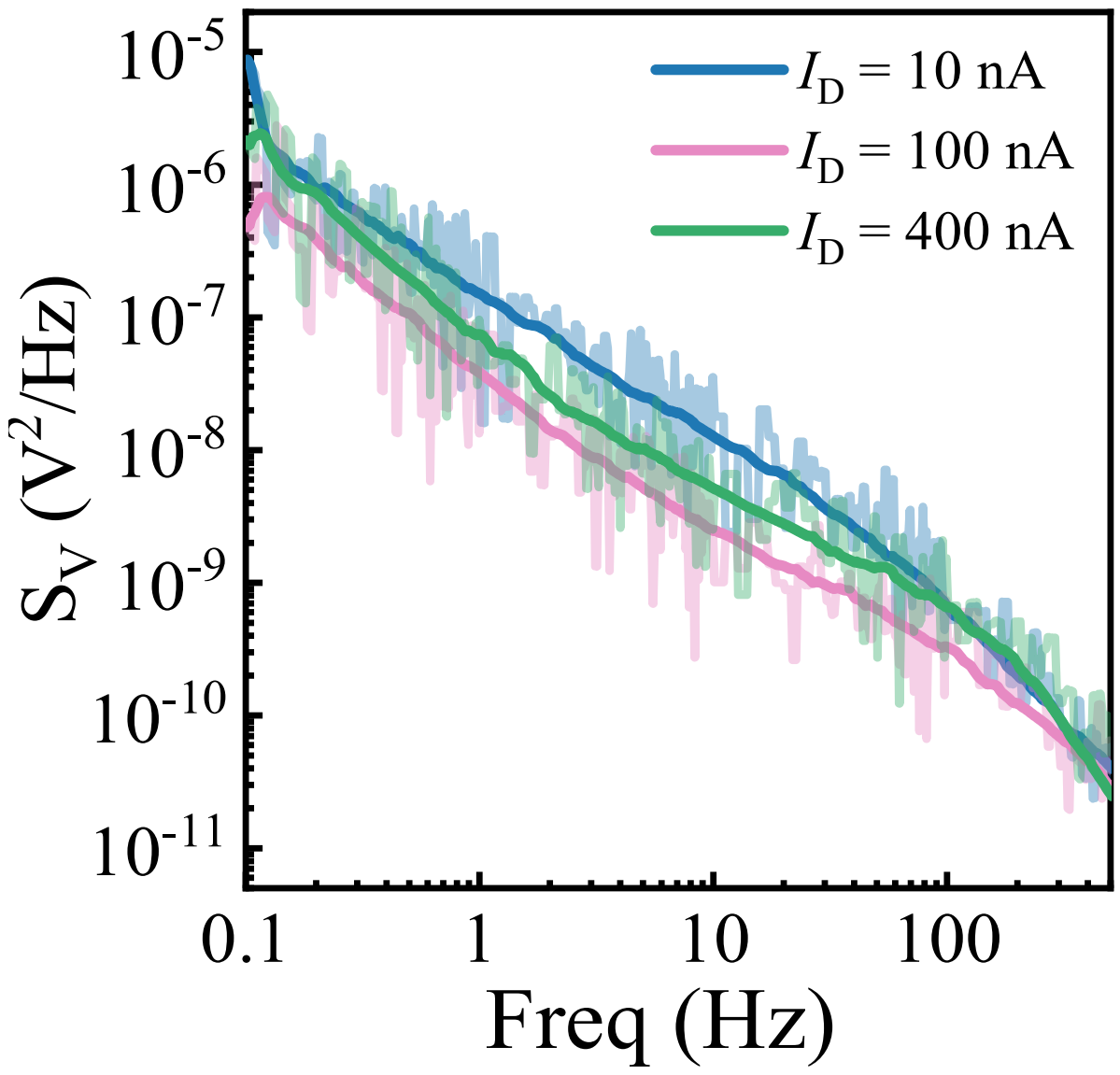

**Fig. S5. $V_G$-referred noise power spectral density ($S_{VG}$) of the SiNWFET.** $S_{VG}$ were measured at different drain currents.

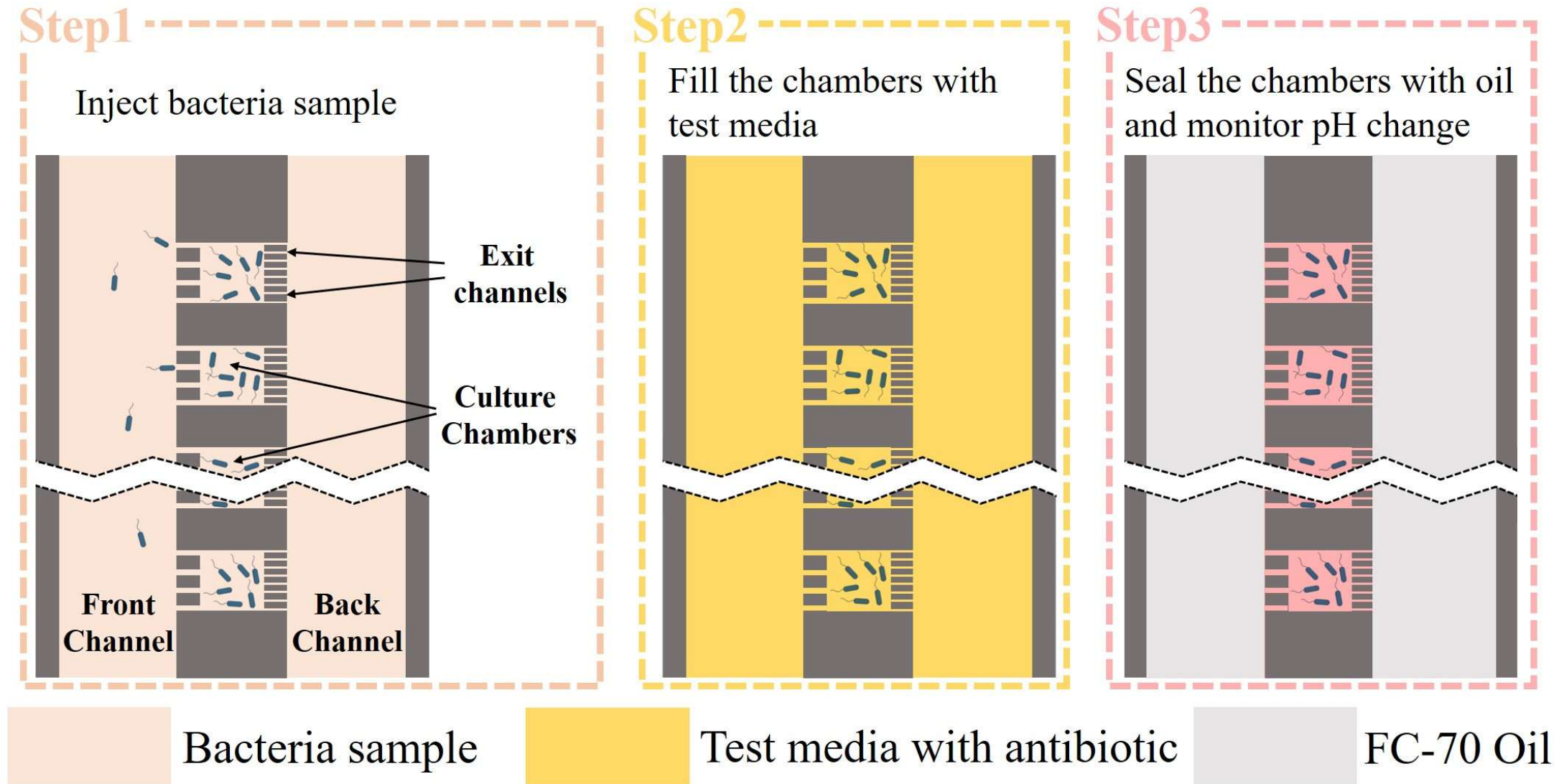

**Fig. S6 Workflow of the AST.** Before AST, LOSC devices were soaked in water at 40 °C for one hour to allow water diffusion into the PDMS layer, reducing evaporation during testing. **Step 1**: Loading of bacterial culture through the front channel; **Step 2**: Introduction of test media (with or without antibiotics); **Step 3**: Sealing of the culture chambers with FC-70 oil, followed by real-time pH monitoring.

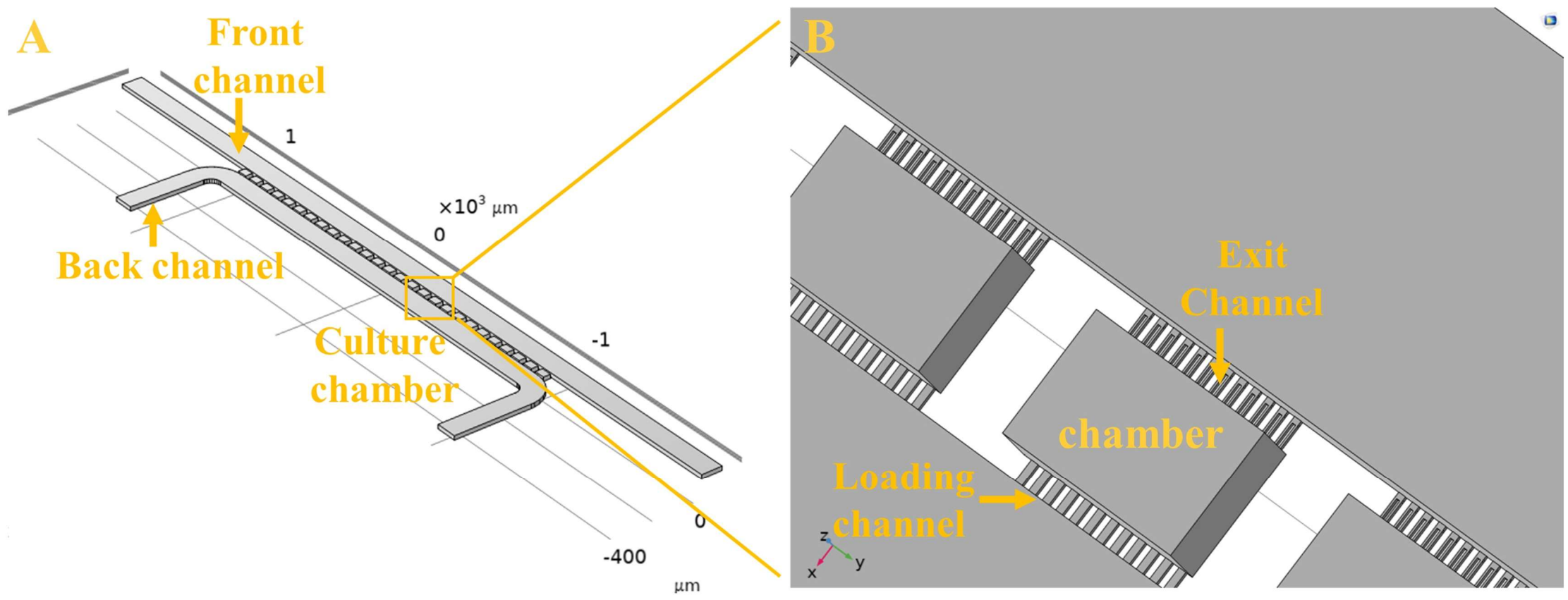

**Fig. S7. 3D model for numerical simulations.** (**A**) 3D model of the cell-collection microfluidics and (**B**) a zoomed-in image of the culture chamber region.

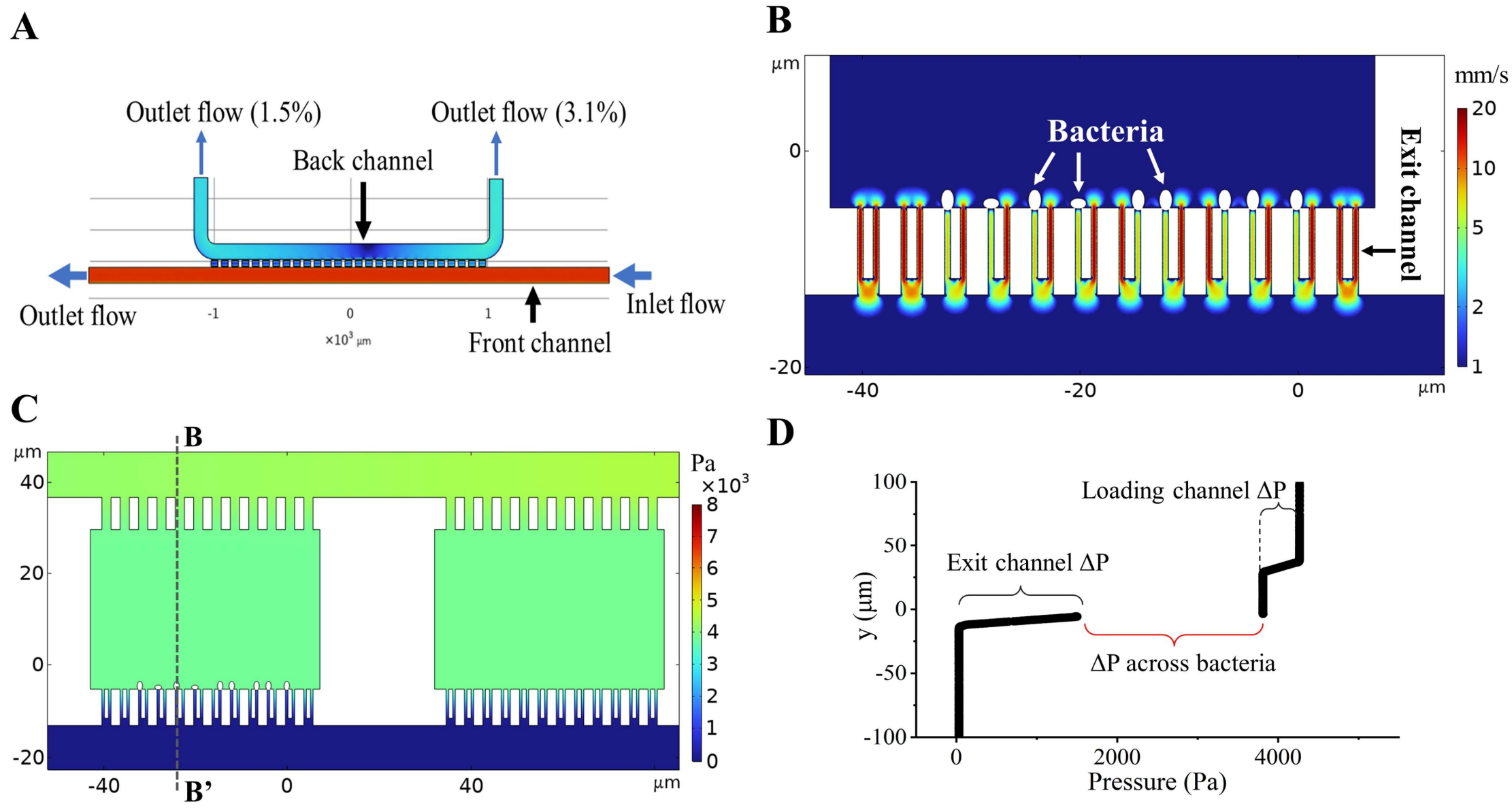

**Fig. S8. Numerical simulation of the flow and pressure profile in the cell-collection microfluidic device under bypass-loading operation.** (**A**) Flow velocity distribution extracted from a vertical plane 0.5 μm above the bottom of microfluidics, showing high velocities in the front channel and reduced velocities inside the chambers and the back channel. Approximately 4.6% of the total flow passes through the culture chambers into the back channel. (**B**) Velocity distribution within the culture chamber. When a bacterium becomes trapped at the entrance of an exit channel, it locally increases the hydraulic resistance of that exit channel, leading to a reduced fluid flux through the occupied chamber. As a result, the incoming flow is redistributed toward less occupied chambers, establishing a self-regulated flow-partitioning mechanism that promotes uniform bacterial loading across chambers. (**C**) Pressure field distribution with bacteria trapped at the exit channel entrances. (**D**) Pressure profile along the dashed line B-B′ in (C), showing pressure drops of ~450 Pa across the loading channel, ~1500 Pa across the exit channel, and ~2300 across the trapped bacterium.

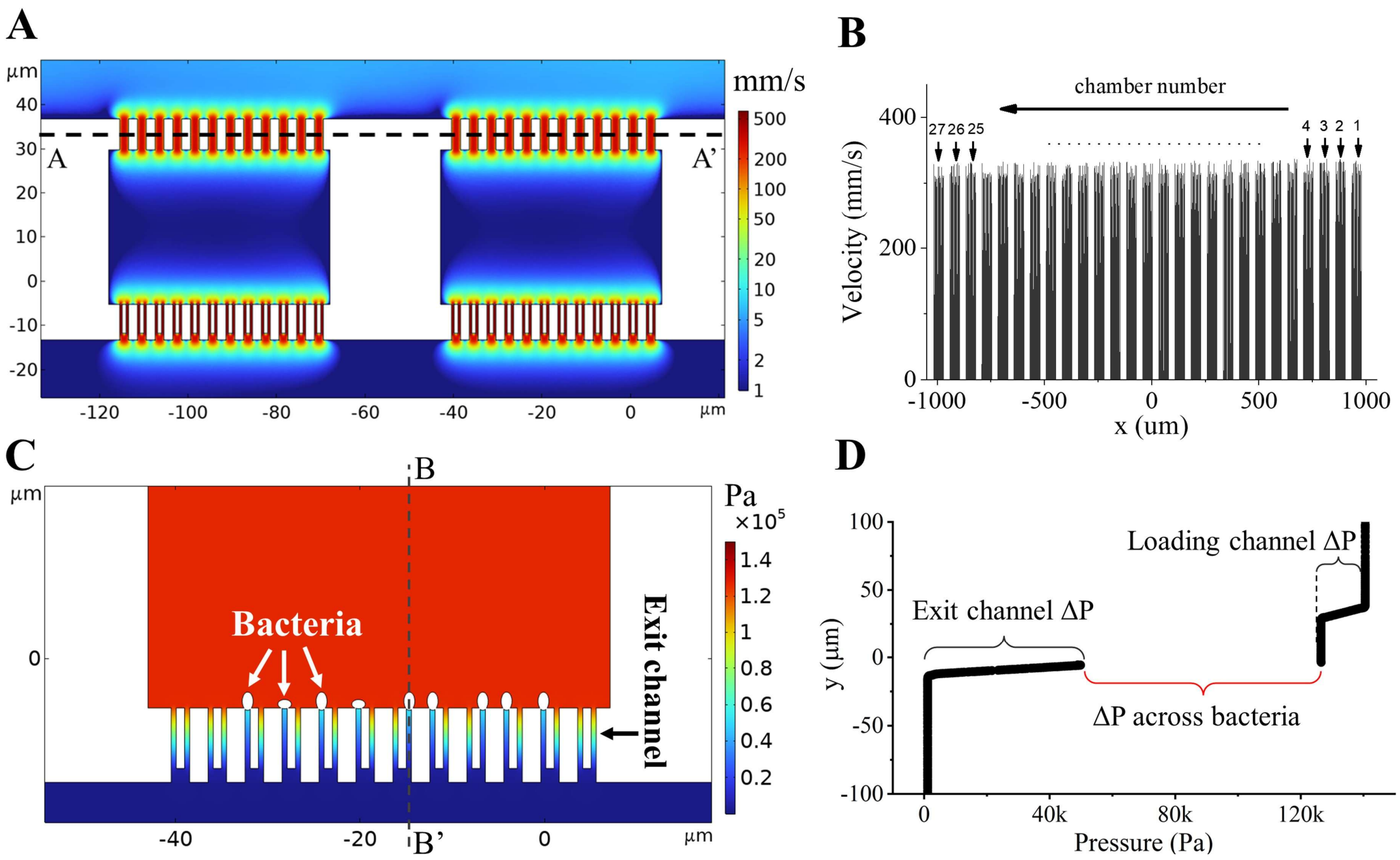

**Fig. S9. Numerical simulation of the flow and pressure profiles in the cell-collection microfluidic device under force-loading operation.** (**A**) Flow velocity distribution extracted from a vertical plane 0.5 µm above the bottom surface, showing high velocities in the loading and exit channels. (**B**) Velocity profile along the dashed line A-A′ in (A), demonstrating nearly identical velocity across all culture chambers, indicating that flow is evenly distributed to each chamber under force-loading conditions. (**C**) Pressure field distribution with bacteria trapped at the entrance of the exit channels. (**D**) Pressure profile along the dashed line B-B′ in (C), showing a pressure drop of ~15 kPa across the loading channel, ~50 kPa across the exit channel, and ~75 kPa across the trapped bacterium.

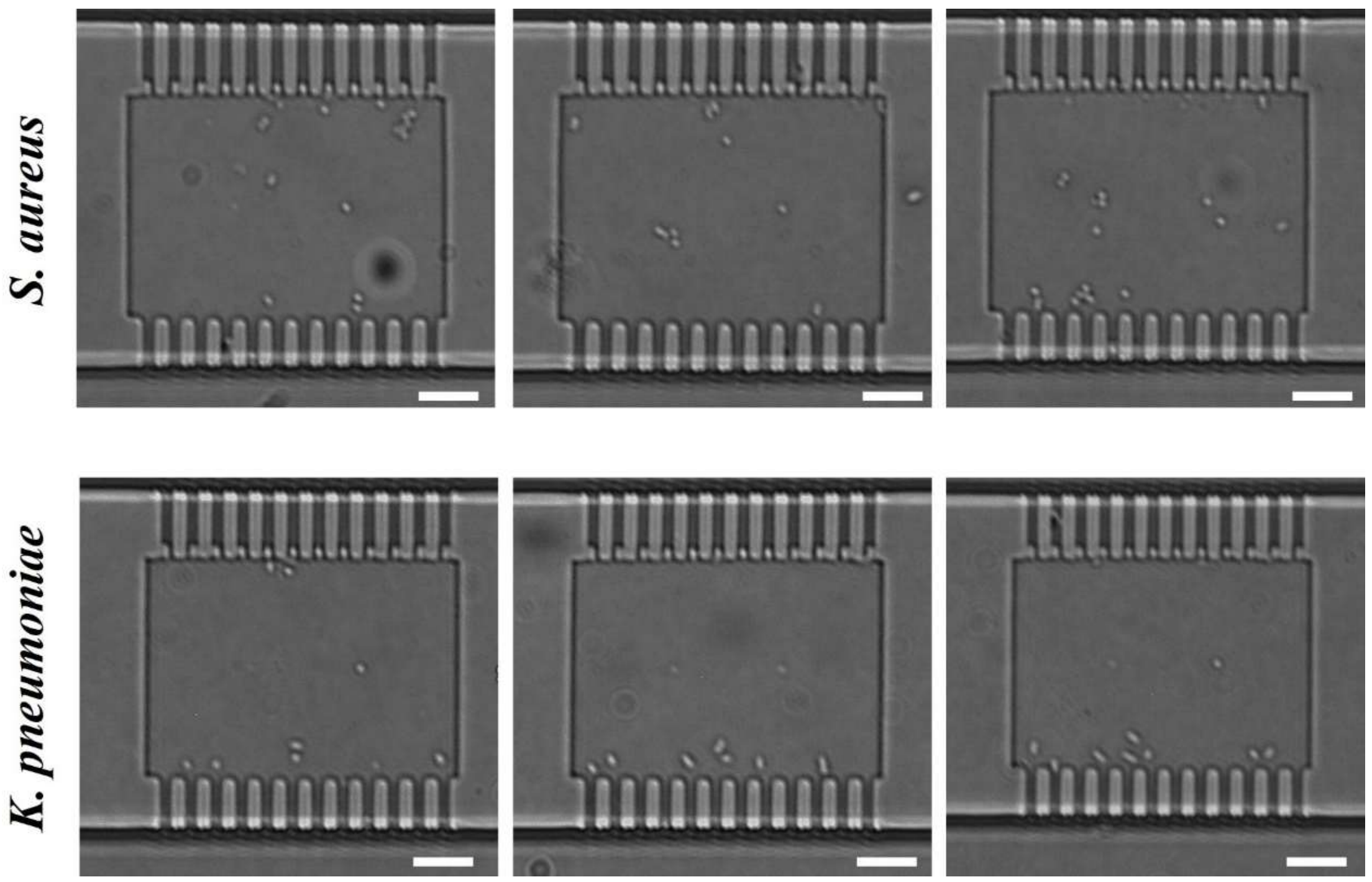

**Figure S10. Loading of *S. aureus* and *K. pneumoniae* into the microchambers.** Images of *S. aureus* and *K. pneumoniae* in microchambers after 8-minute bypass-loading using samples with $1\times10^6$ cells/mL at 8 μL/min. The white scale bars represent 10 μm.

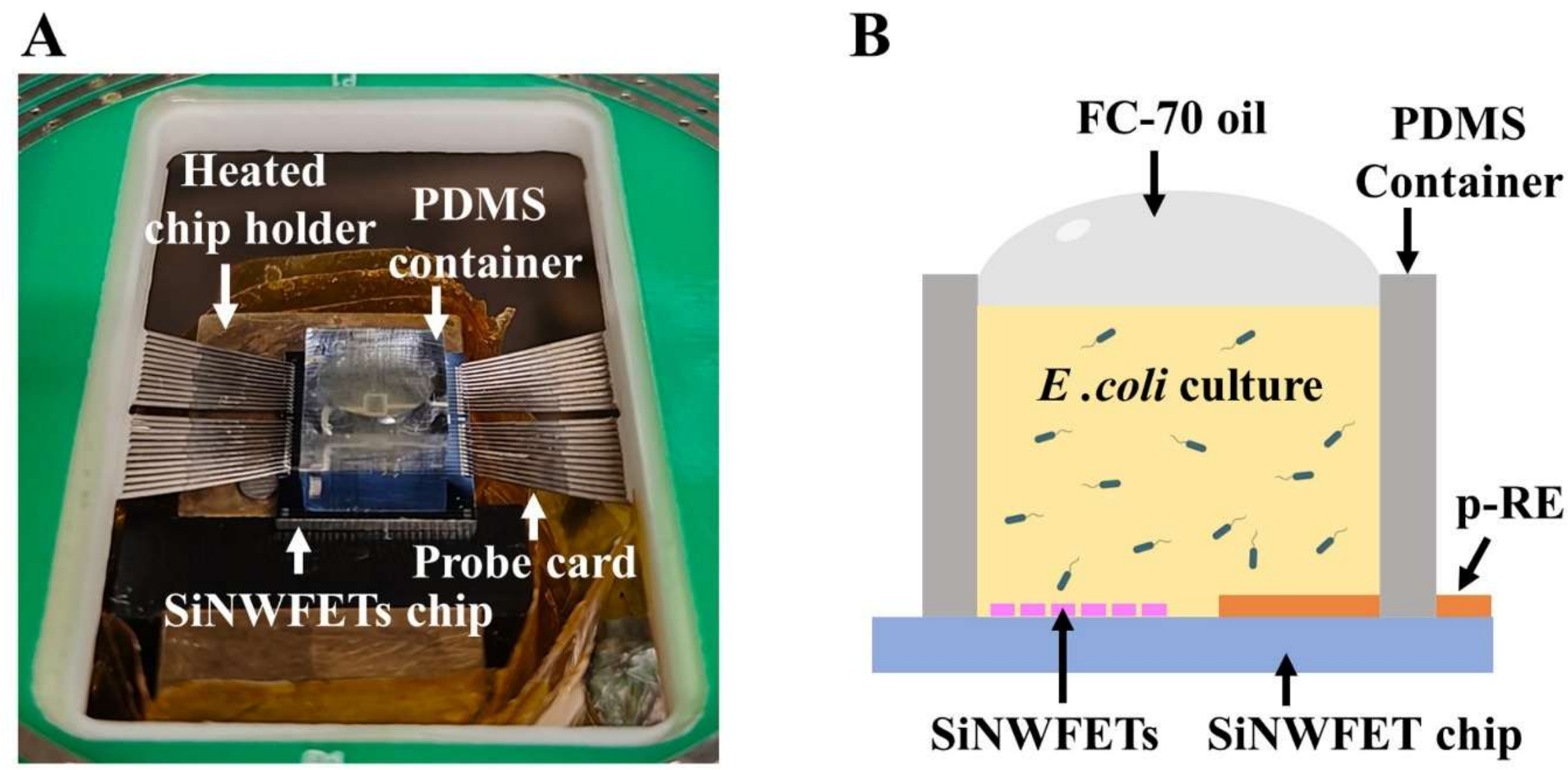

**Figure S11. Bulk cultivation setup.** (**A**) Photograph of the bulk cultivation setup. (**B**) Schematic of the PDMS container filled with *E. coli* sample and sealed with FC-70 oil for metabolism monitoring.

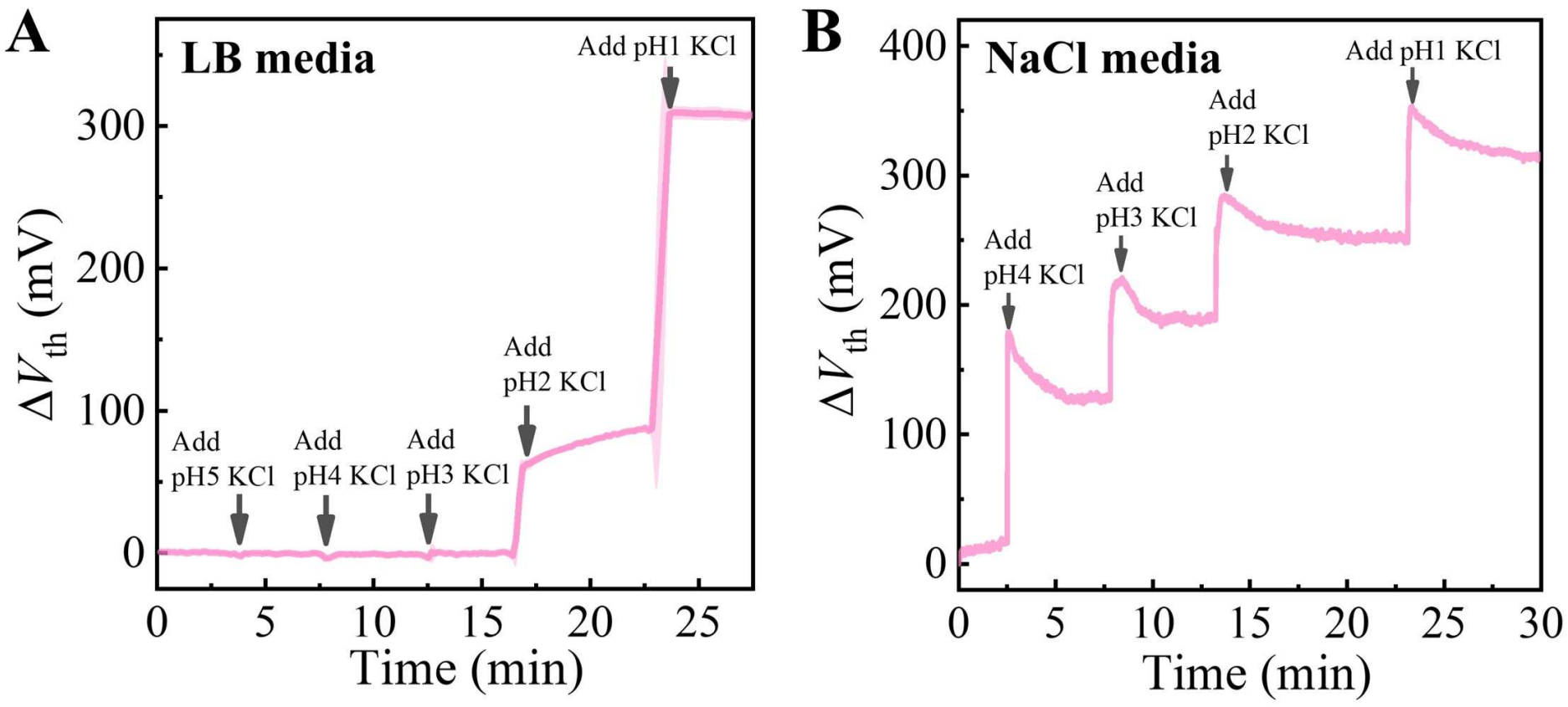

**Fig. S12. Measurement of the pH buffering capacity of different media using SiNWFET.** pH changes were observed when sequentially adding KCl solutions with varying pH to 100 μL of (**A**) LB solution and (**B**) 0.9 % NaCl solution (initial pH is 7).

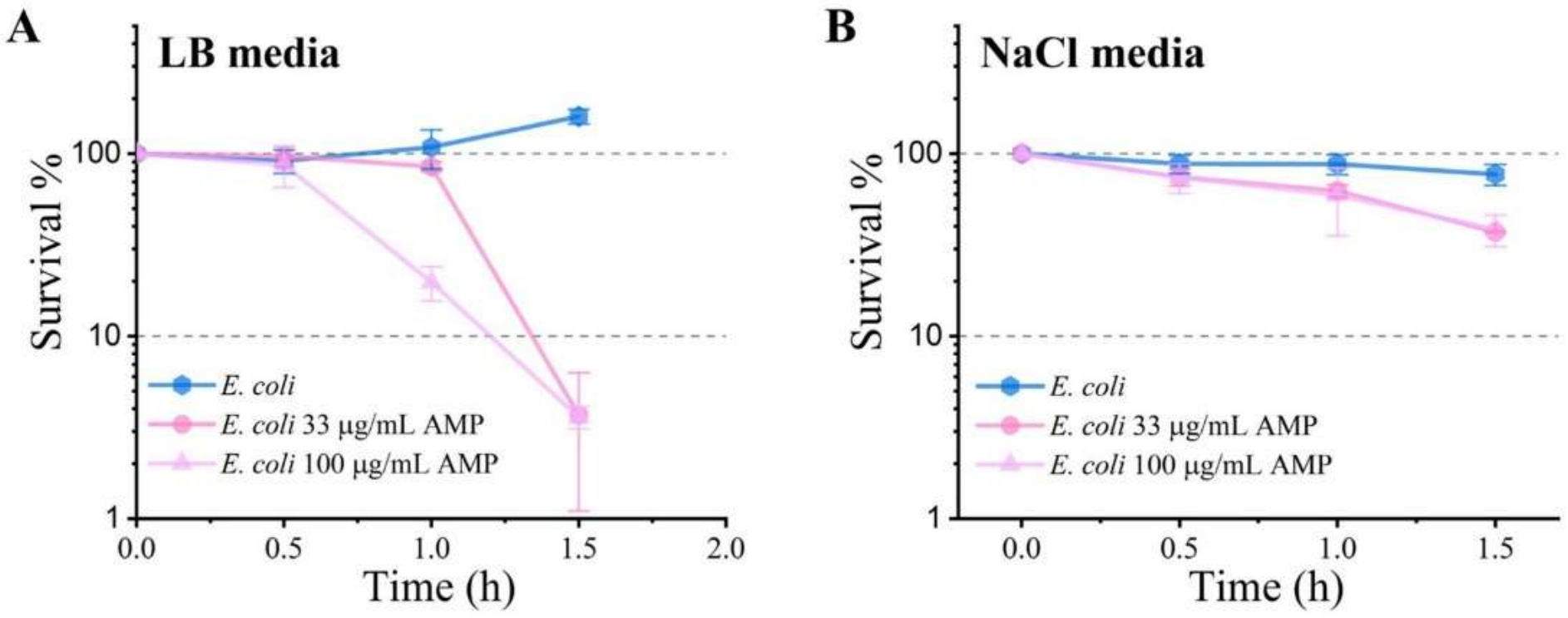

**Fig. S13 Time-kill for *E. coli* with ampicillin treatments.** Overnight cultures of *E. coli* were resuspended in either fresh LB 1% glucose or 0.9% NaCl 1% glucose, with and without AMP at different final concentrations. Survival of *E. coli* after exposure to different concentrations of AMP was calculated. (**A**) The response of *E. coli* to AMP in LB media shows a reduction in survival. (**B**) The response of *E. coli* to AMP in NaCl media shows some reduction in survival. Error bars show the standard error, N=3 biological replicates.

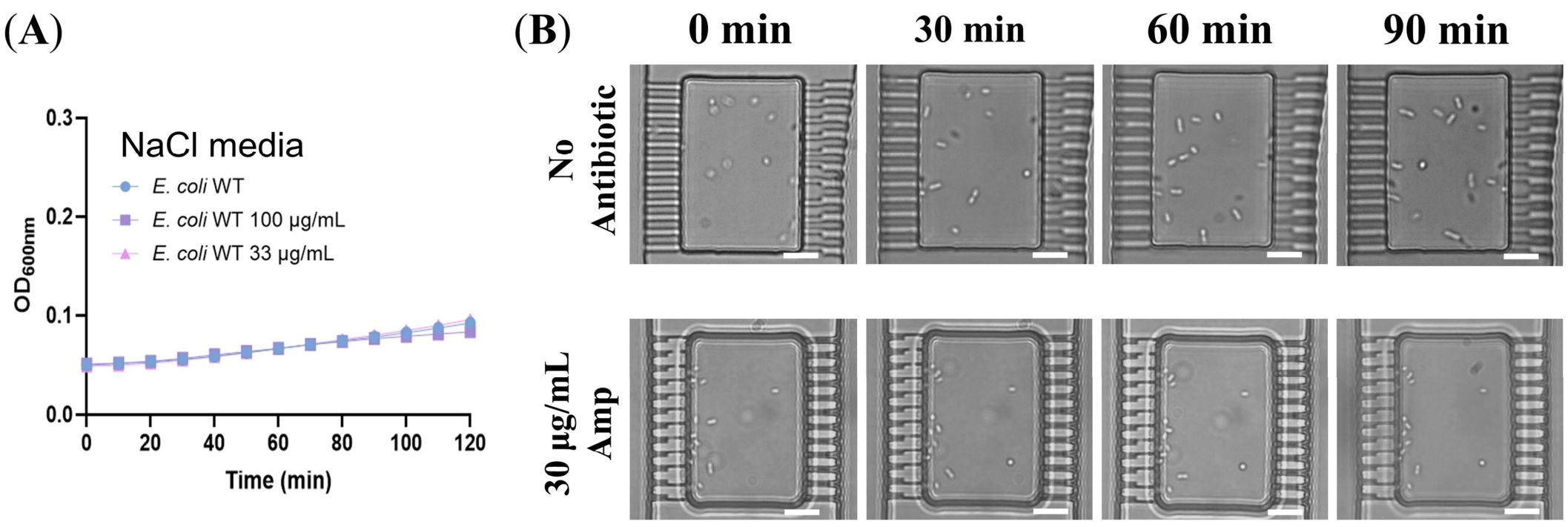

**Fig. S14 Growth of *E. coli* under ampicillin treatment.** (**A**) $OD_{600}$ of *E. coli* cultured in NaCl medium in the presence of AMP. (**B**) Time-lapse images showing *E. coli* growth in microchambers filled with NaCl medium without antibiotics (top row) and with AMP (bottom row). The white scale bars represent 10 μm.

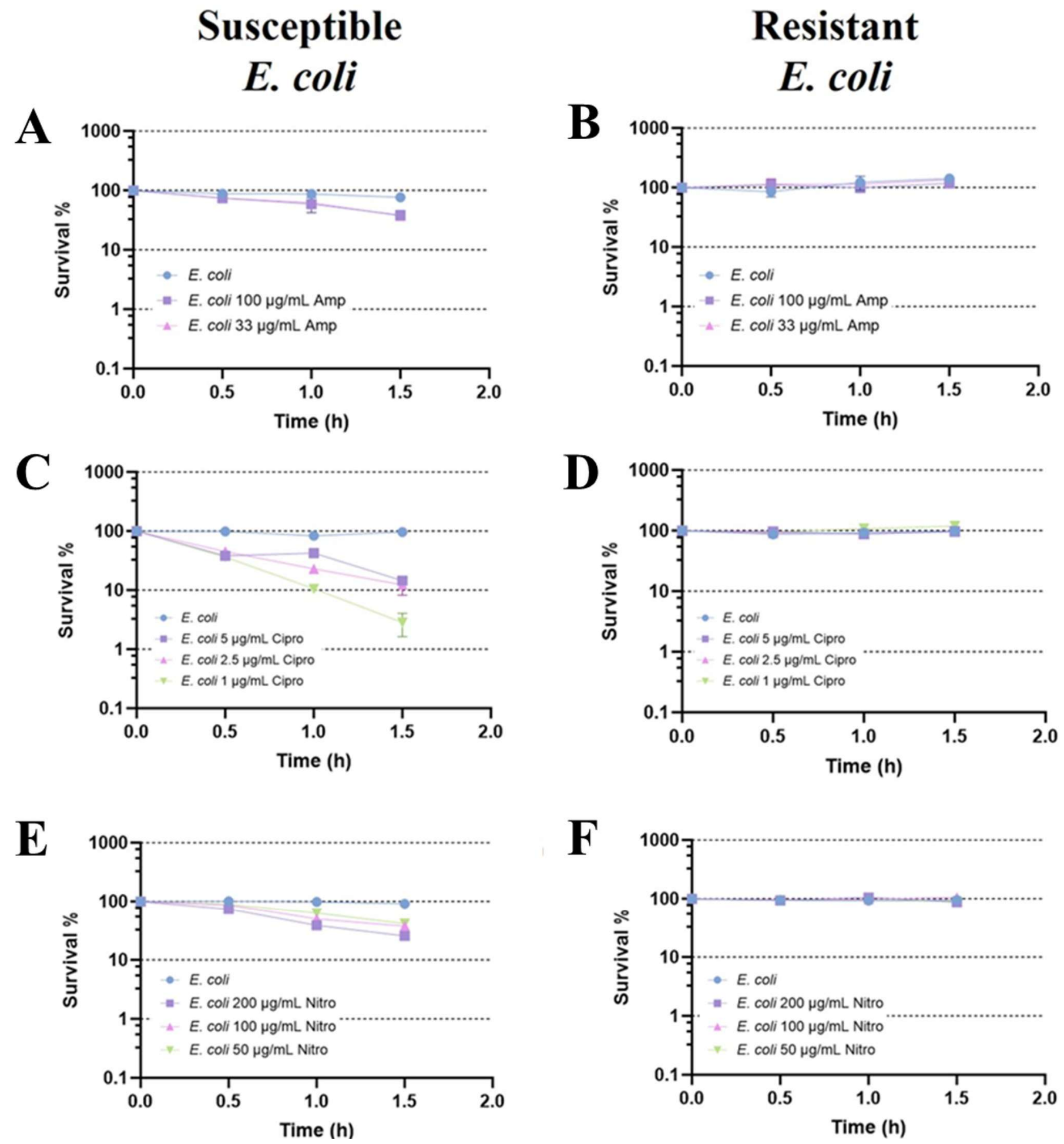

**Fig. S15 Time-kill for *E. coli* exposed to different antibiotics in 0.9% NaCl with 1% glucose.** Survival of susceptible (**A**, **C**, **E**) and resistant (**B**, **D**, **F**) *E. coli* after exposure to (**A** and **B**) AMP (33 and 100 μg/mL), (**C** and **D**) CIP (1, 2.5, and 5 μg/mL), and (**E** and **F**) NIT (50, 100, and 200 μg/mL), respectively. Error bars are the standard error. N=3 biological replicates.

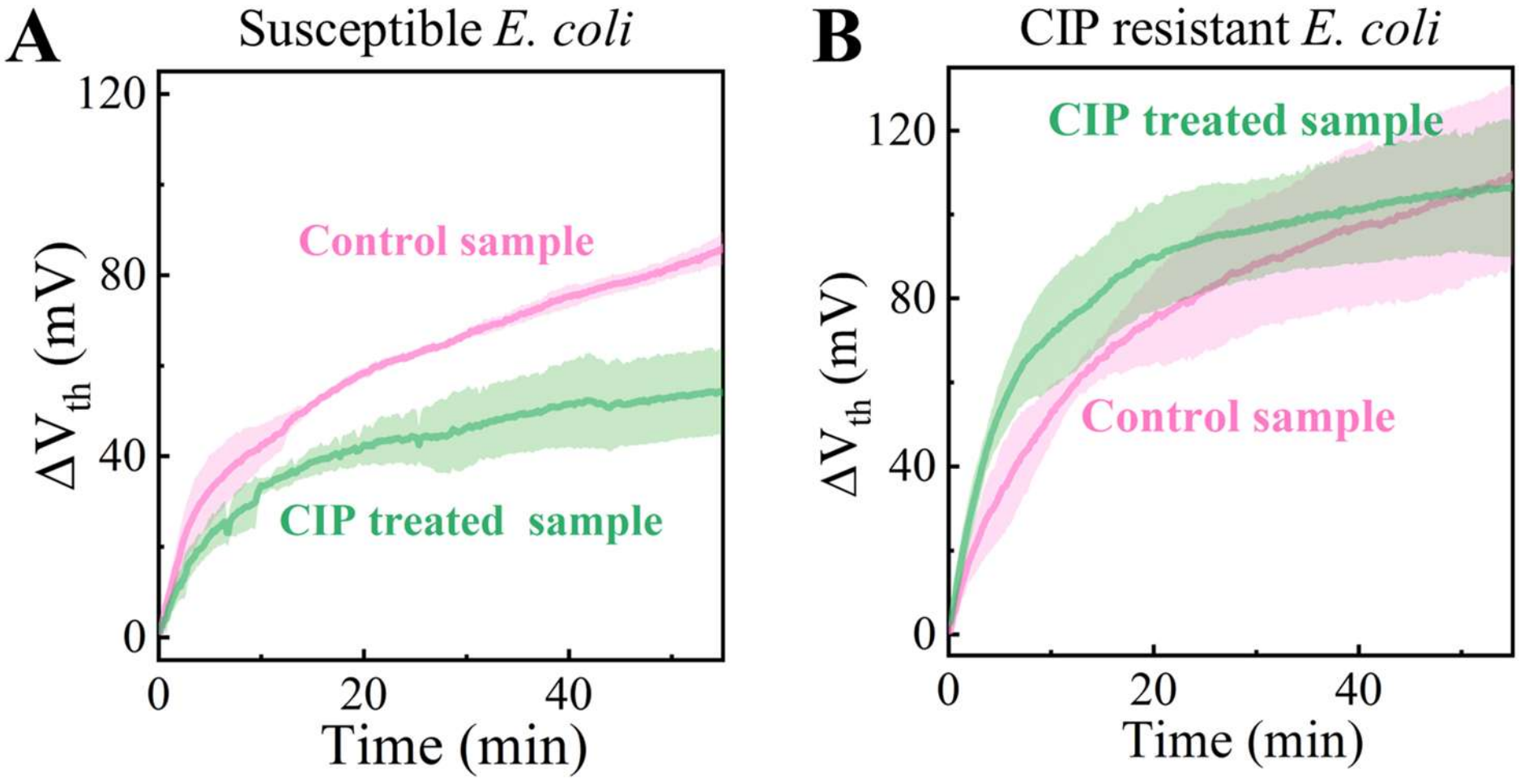

**Fig. S16** The LOSC AST with three biological replicates for (**A**) susceptible *E. coli* and (**B**) CIP-resistant *E. coli* (medium resistant).

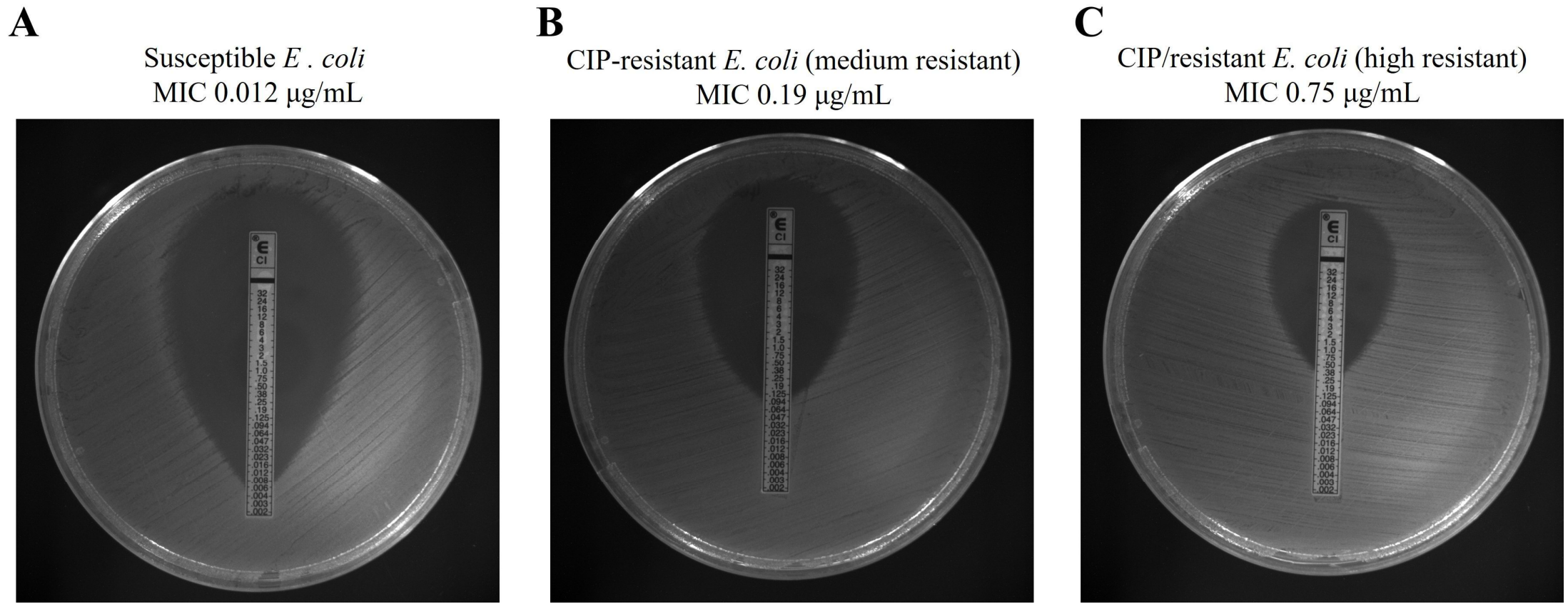

**Fig. S17. E-test determination of CIP minimum inhibitory concentration (MIC) for three *E. coli* strains.** (**A**) Susceptible *E. coli*, MIC = 0.012 μg/mL; (**B**) CIP-resistant *E. coli* (medium resistance), MIC = 0.19 μg/mL; (**C**) CIP-resistant *E. coli* (high resistant), MIC = 0.75 μg/mL.

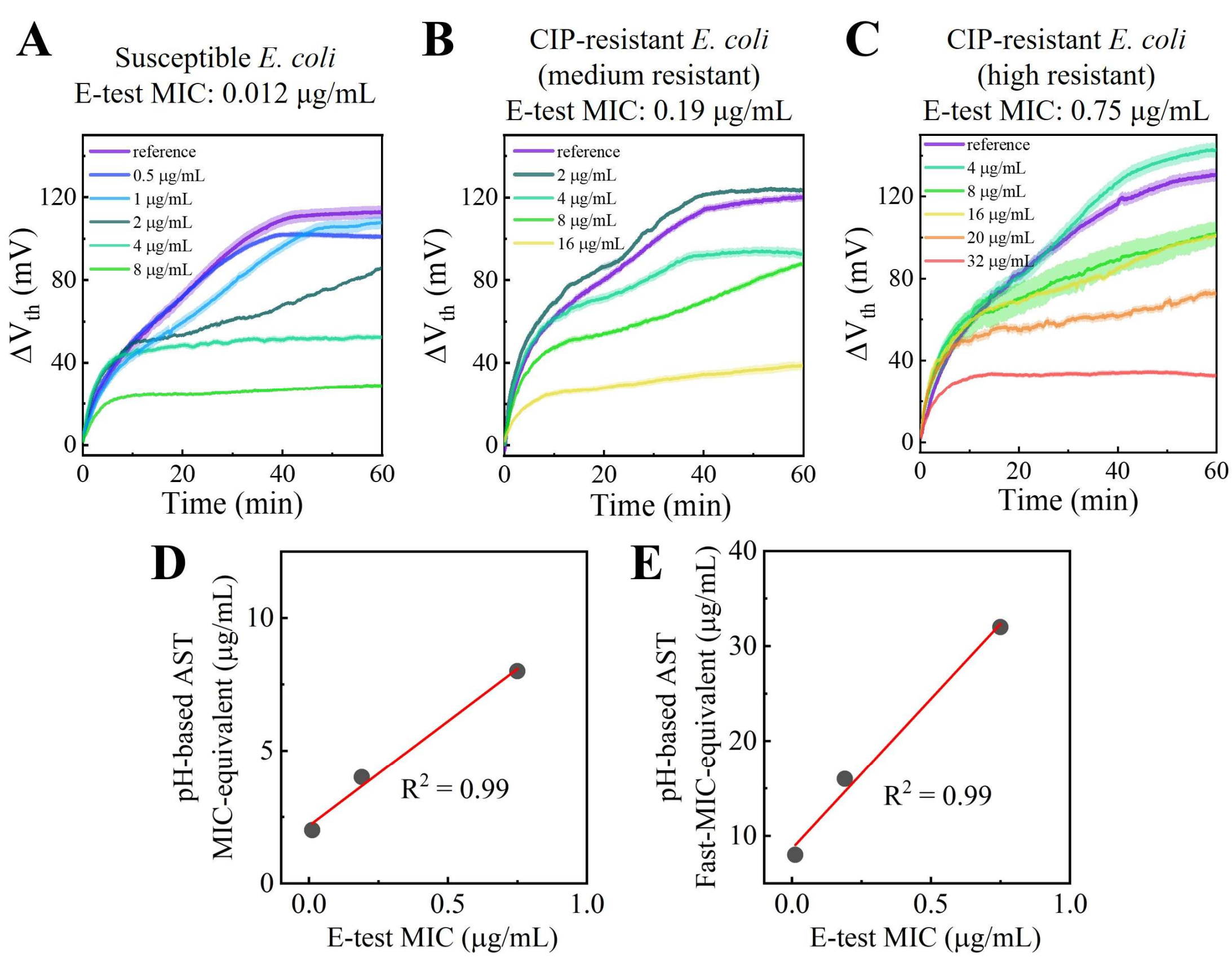

**Fig. S18. Alignment of pH-based AST with standard MIC measurements.** (**A-C**) $\Delta V_{th}$ responses of three *E. coli* strains to various CIP concentrations, with E-test MICs of (**A**) 0.012, (**B**) 0.19, and (**C**) 0.75 μg/mL, measured at ~$3 \times 10^8$ cells/mL in 0.9% NaCl with 1% glucose. (**D**) MIC-equivalent values extracted using a general criterion ($\Delta V_{th}$ difference > 20 mV during the measurement period), yielding 2, 4, and 8 μg/mL, showing linear correlation with E-test MICs ($R^2 = 0.99$). (**E**) FAST-MIC-equivalent values extracted using a rapid criterion ($\Delta V_{th}$ difference > 20 mV within 5 min), yielding 8, 16, and 32 μg/mL, showing linear correlation with E-test MICs ($R^2 = 0.99$).

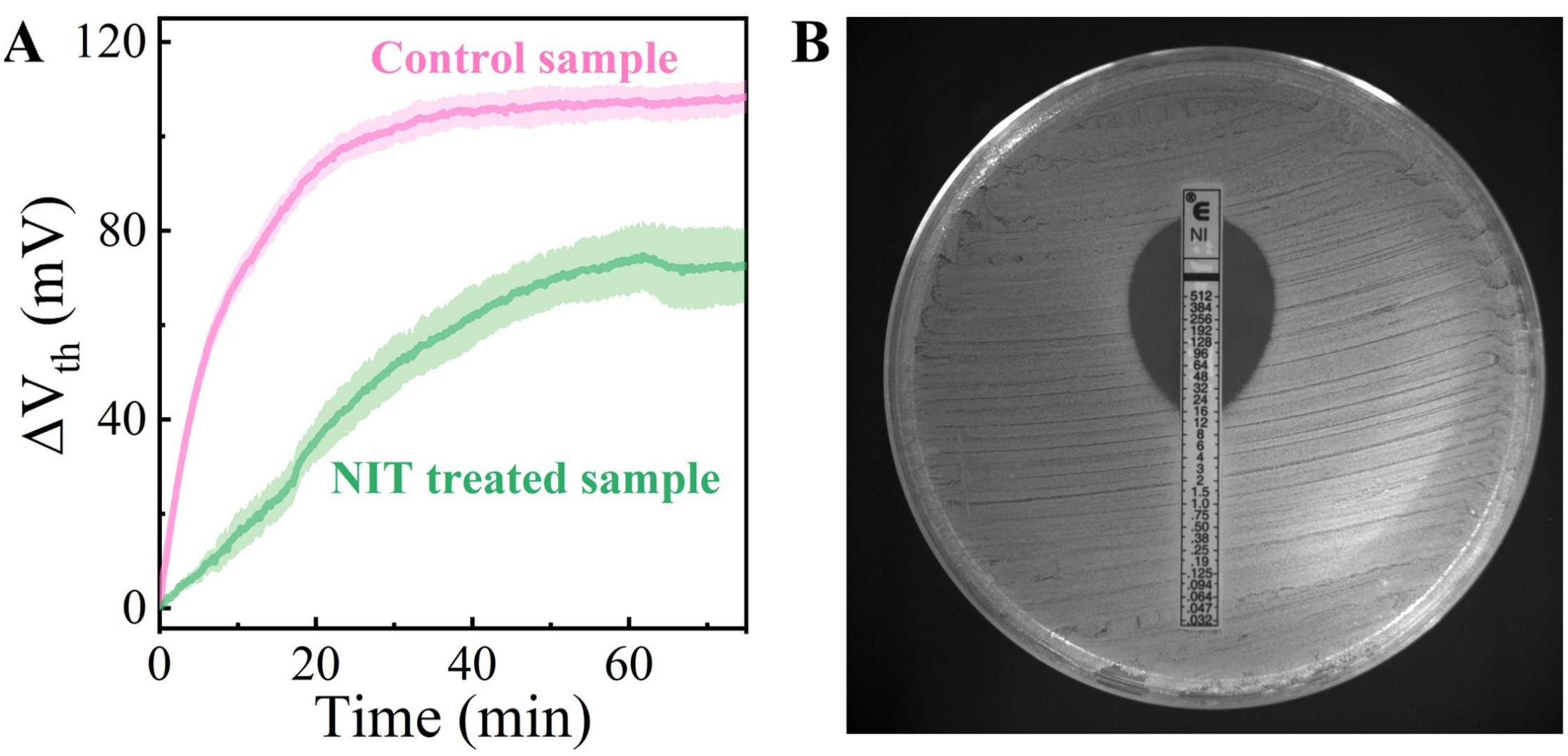

**Fig. S19 ASTs of *S. aureus* using the LOSC platform in 0.9 wt.% NaCl (1 wt. % glucose).** (**A**) $\Delta V_{th}$ vs. time for *S. aureus* in response to NIT (200 µg/mL). During AST, *S. aureus* samples ($1 \times 10^6$ cells/mL) were loaded into the LOSC device at 8 µL/min for 8 minutes using the bypass mode. (**B**) E-test for *S. aureus* against NIT, confirming that the *S. aureus* strain (MIC of 16 µg/mL) is susceptible to NIT (*73*).

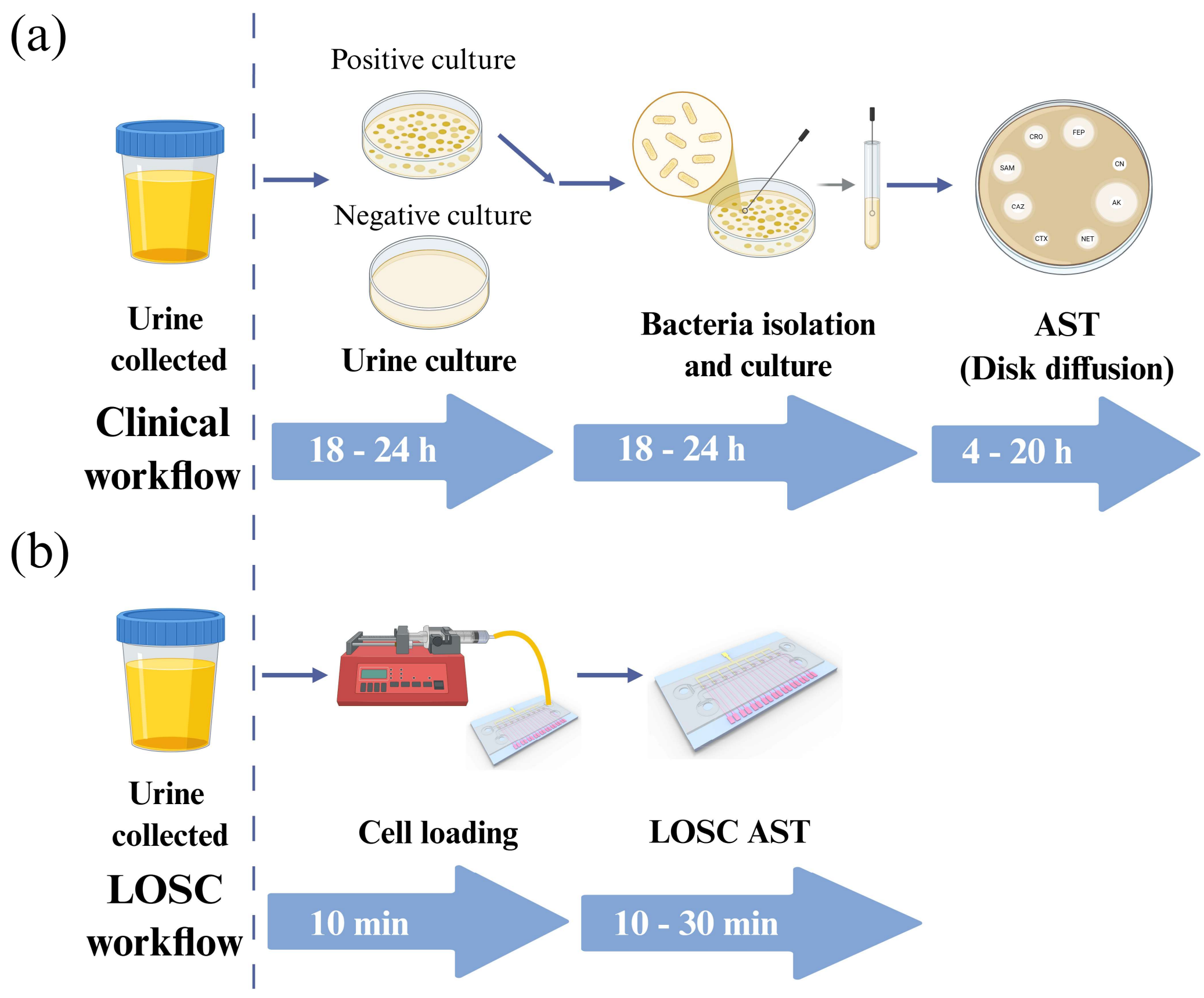

**Fig. S20 Sample-to-result time of different AST workflows.** Timeline comparing the (**A**) current clinical workflow to the (**B**) LOSC workflow. Created in BioRender. Xu, Z. (2026) https://BioRender.com/1hr0kki

# Supplementary tables

**Table. S1** ANOVA analysis of bacterial loading uniformity across chamber positions

| Cell density of the Loading sample | Chamber group1 | | Chamber group2 | | Chamber group3 | | p |
|---|---|---|---|---|---|---|---|
| | Mean | Variance | Mean | Variance | Mean | Variance | |
| $5\times10^{6}$ cell/mL [1] | 8.3 | 3.0 | 8.4 | 2.1 | 7.8 | 3.7 | 0.49 |
| $5\times10^{4}$ cell/mL [2] | 4.4 | 1.8 | 4.1 | 1.5 | 4.5 | 1.5 | 0.47 |
| $1\times10^{3}$ cell/mL [3] | 4.1 | 1.1 | 4.9 | 2.2 | 5.1 | 1.5 | 0.15 |

[1] An *E. coli* sample with a concentration of $5\times10^{6}$ cells/mL was introduced into the cell-collection microfluidics using bypass loading mode at a flow rate of 8 μL/min for 8 minutes.
[2] An *E. coli* sample with a concentration of $5 \times 10^{4}$ cells/mL was introduced into the cell-collection microfluidics using force loading mode at a flow rate of 8 μL/min for 8 minutes.
[3] An *E. coli* sample with a concentration of $1 \times 10^{3}$ cells/mL was introduced into the cell-collection microfluidics using force loading mode at a flow rate of 8 μL/min for 16 minutes.

**Table. S2** Genotype of the bacterial strain used in this study

| **Strain** | **Sample name** | **Genotype** |
|---|---|---|
| SK2 | Susceptible *E. coli* | *E. coli* K12 MG1655 |
| SK4582 | AMP-resistant *E. coli* | *E. coli* K12 MG1655 *lacIZYA::bla* |
| SK4686 | CIP-resistant *E. coli* (medium resistant) | *E. coli* K12 MG1655 *gyrA*(D87N) |
| SK8157 | CIP-resistant *E. coli (high resistant)* | DA25191, *gyrA*(S83L) *parC*(S801) |
| SK7394 | NIT-resistant *E. coli* | *E. coli K*12 MG1655 D*nsfAB* |
| SK161 | UPEC | *Uropathogenic E. coli* 536 |
| SK5200 | Susceptible *S. aureus* | *S. aureus* ATCC BAA-1556 |
| SK1314 | *K. pneumoniae* | *K. pneumoniae* (ATCC 13883/NCTC 9633) |